\documentclass[]{aastex631}

\newcommand{\sk}[1]{\textcolor{black}{#1}}
\newcommand{\COM}[1]{\textcolor{black}{#1}}
\newcommand{\COMs}[1]{\textcolor{black}{#1}}
\usepackage[normalem]{ulem}
\usepackage{cancel}
\usepackage{microtype}
\accepted{\today}

\begin{document}

\title{Similar image retrieval using Autoencoder. I. Automatic morphology classification of galaxies}

\correspondingauthor{Suk Kim}
\email{star4citizen@gmail.com}

\author{Eunsuk Seo}
\affiliation{Department of Astronomy, Space Science and Geology, Chungnam National University, Daejeon 34134, Republic of Korea}

\author[0000-0003-3474-9047]{Suk Kim}
\affiliation{Department of Astronomy and Space Science \& Research Institute of Natural Sciences, Chungnam National University, Daejeon 34134, Republic of Korea; star4citizen@gmail.com}

\author[0000-0002-6261-1531]{Youngdae Lee}
\affiliation{Department of Astronomy and Space Science \& Research Institute of Natural Sciences, Chungnam National University, Daejeon 34134, Republic of Korea; star4citizen@gmail.com}

\author{Sang-Il Han} 
\affiliation{Department of Science Education, Ewha Womans University, Seoul 03760, Republic of Korea}

\author[0000-0001-7033-4522]{Hak-Sub Kim} 
\affiliation{Department of Physics and Astronomy, Sejong University, Seoul 05006, Republic of Korea}

\author[0000-0002-0041-6490]{Soo-Chang Rey}
\affiliation{Department of Astronomy and Space Science, Chungnam National University, Daejeon 34134, Republic of Korea}

\author{Hyunmi Song}
\affiliation{Department of Astronomy and Space Science, Chungnam National University, Daejeon 34134, Republic of Korea}



\begin{abstract}
 \COM{We present the construction of an image similarity retrieval engine for the morphological classification of galaxies using the Convolutional AutoEncoder (CAE). The CAE is trained on 90,370 preprocessed Sloan Digital Sky Survey galaxy images listed in the Galaxy Zoo 2 (GZ2) catalog. The visually similar output images returned by the trained CAE suggest that the encoder efficiently compresses input images into latent features, which are then used to calculate similarity parameters. Our Tool for Searching a similar Galaxy Image based on a Convolutional Autoencoder using Similarity (TSGICAS) leverages this similarity parameter to classify galaxies' morphological types, enabling the identification of a wider range of classes with high accuracy compared to traditional supervised ML techniques. This approach streamlines the researcher's work by allowing quick prioritization of the most relevant images from the latent feature database. We investigate the accuracy of our automatic morphological classifications using three galaxy catalogs: GZ2, Extraction de Formes Idéalisées de Galaxies en Imagerie (EFIGI), and Nair $\&$ Abraham (NA10). The correlation coefficients between the morphological types of input and retrieved galaxy images were found to be 0.735, 0.811, and 0.815 for GZ2, EFIGI, and NA10 catalogs, respectively. Despite differences in morphology tags between input and retrieved galaxy images, visual inspection showed that the two galaxies were very similar, highlighting TSGICAS's superior performance in image similarity search. We propose that morphological classifications of galaxies using TSGICAS are fast and efficient, making it a valuable tool for detailed galaxy morphological classifications in other imaging surveys.}

\end{abstract}



\section{Introduction} \label{sec:intro}
Galaxy morphologies can provide critical information regarding galaxy research because galaxy morphology is related to galaxy properties, evolutionary history, dynamical features, and stellar populations. The galaxies were first classified morphologically by \cite{Hubble1926} via visual assessments. Subsequently, \cite{Vaucouleurs1959} divided the Hubble system’s morphological classes and proposed a modified Hubble system. In recent decades, the size of galaxy image datasets has increased considerably with the releases of large sky survey projects, such as the Sloan Digital Sky Survey \citep[SDSS;][]{York2000}. With the development of computational power, nonparametric galaxy morphological classification methods, such as the concentration, asymmetry, smoothness/clumpiness systems, Gini coefficients, and M20, were developed \citep{Abraham2003,Conselice2003,Lotz2004,Law2007}. These methods are free from subjective bias, but they have limitations in detail classifying galaxy morphologies such as the Hubble system. Therefore, visual classification is still employed in most galaxy morphological classification. 

Galaxy Zoo 2 (GZ2) project \citep{Lintott2008,Lintott2011,Willett2013} is representative citizen science, wherein amateurs quickly classify galaxies by answering a series of galaxy image-related questions. GZ2 successfully analyzed large-scale galaxy morphology data.
Despite the successful implementation of the GZ2 project, the sheer volume of data from potential large-scale astronomical surveys (\COM{The Rubin Observatory LSST Camera (LSSTCam)}, James Webb Space Telescope (JWST), and Square Kilometer Array (SKA)) may significantly extend the time required for the classification spanning decades or even centuries, and necessitate the use of alternative classification methods. The classification method using Machine Learning (ML) is one of the crucial alternatives in large astronomical dataset analysis.
Therefore, the use of ML techniques in large astronomical dataset analysis is crucial. 

ML-based computational tools were introduced by \cite{Fukushima1975,Fukushima1980} and \citet*{Fukushima1983}. ML was developed to suit multiple applications, including the earlier astronomical investigations (\citealt{Odewahn1992}; \citealt*{Weir1995}). The enhancement of computational power and ML methodologies has led to the emergence of ML as an important astronomical research tool. Supervised ML(SML) approaches have been applied in galaxy morphological classification \citep{Huertas-Company2008,Huertas-Company2009,Huertas-Company2011,Shamir2009,Dubath2011,Polsterer2012,Beck2018,Sreejith2018,Maehoenen1995,Naim1995,Lahav1996,Ball2004,Dieleman2015,Huertas-Company2015,Huertas-Company2018,Cheng2020A,Walmsley2020}. 

Convolutional Neural Networks (CNNs) have been utilized in image-related studies because of their unique ability to store image location information in their layers, in contrast to the traditional Multi-layer Perceptron (MLP) approach.
Using ImageNet data in CNN-based models \citep{Deng09} has shown higher accuracy than traditional manual classification tools.
Therefore, CNNs have been employed for galaxy morphological classification. 
CNN-based SML relies heavily on the availability of labeled datasets, as it mimics human perception by training models with human-labeled data. It can accurately classify the galaxies
\citep{DeLaCalleja2004,Khalifa2017,Cavanagh2021,Cheng2021}. However, the labeled datasets with sufficient quantity and quality are required to apply them to future surveys \COM{\citep{Cheng2023,Vega-Ferrero2021,Ghosh2020}}.  

One of the critical advantages of Unsupervised Machine Learning (UML) approaches is that they do not rely on labeled training datasets, making the acquisition of training data much more straightforward. These techniques have been successfully applied to classify celestial bodies such as galaxies by extracting features such as spiral arm patterns, bulges, and disk shapes \COM{\citep{Cheng2021a,Cheng2020, Ralph2019,Hocking2018}}. This method allows the grouping of galaxies with similar morphological characteristics, making it easier to label their morphology.


With the remarkable development of ML, a novel technique has been developed for effectively compressing image features used for similar image searches in the industry (e.g., face recognition on Facebook). 
An autoencoder, as described by \cite{Rumelhart1986}, is a type of UML network utilized for feature extraction from input images. The encoder component employs convolutional layers to generate latent features, which are then reconstructed by the decoder into an output image of the same size. The training continues until a low residual between the input and output images is achieved.
The latent features generated by the encoder encapsulate essential features of the input image, including morphological parameters in the case of galaxy images. Since the proximity of latent features between two images indicates their similarity, these latent features can be utilized to compare and retrieve similar images effectively.

If a database of latent features is constructed with galaxy images of known morphological types, classification can be performed by adopting the morphological types of similar images retrieved from the database. This method is more effective than conventional morphological parameters (concentration, asymmetry, smoothness/clumpiness systems, Gini coefficients, M20, etc.) because the encoder can identify the image features more effectively. \COM{Similarity search has been applied in various astronomical studies, including galaxy morphology research, such as \citep{Walmsley2022}, which focuses on the Galaxy Zoo projects. Moreover, anomaly detection studies, like \citep{Storey-Fisher2021}, can also be considered as similarity search applications in a broader sense.} As the Convolutional AutoEncoder (CAE) extracted the latent features of galaxy images regardless of the labels, it does not require re-training upon morphological types switch, unlike the SML.

While the conventional morphological system (like the Hubble system) classifies galaxies into more than ten morphological classes, most ML-based system classifies them into four or five morphological types \citep{Cavanagh2021, Gauci2010}. We developed the Tool for Searching similar Galaxy Image based on a Convolutional Autoencoder using Similarity(TSGICAS), which can distinguish more than ten morphological types.


The remainder of this paper is organized as follows.  Section \ref{sec:data} describes the model training datasets and the preprocessing method. Section \ref{sec:method} describes the CAE’s algorithm, training, validation of similar image searching methods, the encoder’s latent feature extraction, and database creation methods. Section \ref{sec:classify} describes morphological classifications for three galaxy catalogs. Section \ref{sec:improve} discusses model improvement. Section \ref{sec:summary} presents a summary and conclusions.  

\section{Data and Preprocessing} \label{sec:data}

\COM{In modern observational astronomy, numerous survey projects have generated extensive open-source data.  This data is primarily machine readable and has been standardized for scientific analysis, making it suitable for machine learning applications. The Sloan Digital Sky Survey (SDSS), a representative astronomical survey, is a large-scale optical survey project that provides the largest and highest-quality photometric and spectroscopic data. Our objective is to classify galaxy morphological types based on the conventional Hubble classification. To achieve this using TSGICAS, it is essential to know the morphological types of retrieved galaxies. Therefore, we adopted galaxy morphologies from three morphological catalogs: GZ2, Extraction de Formes Idealisees de Galaxies en Imagerie \citep[EFIGI;][]{Baillard2011} and NA10 \citep{Nair2010}. In the GZ2 catalog, a representative citizen science project, the galaxy morphologies are classified by a statistical method to minimize the amateur classifier's subjective influence. The EFIGI and NA10 catalogs were classified by a small number of expert astronomers via their own subjective classification but exhibited consistent classifications}

\COM{\subsection{Morphology Catalogs and Training data}}

The GZ2 morphological classification decision tree was created using the “tuning fork” approach \citep{Hubble1926}. The decision tree contained 11 questions and 37 responses. The classification followed the question order and the highest vote fractions. The smooth elliptical galaxies (E) were categorized into Er (round), Ei (in between), and Ec (cigar-shaped). The spiral galaxies with an arm structure or disk shape were divided into spiral galaxies with bars (SB) and without (S), respectively. The SB and S were further classified into a (dominant), b (obvious), c (just noticeable), and d (no bulge), following their size of the bulge. Their spiral arm and winding patterns were not considered. We reduced the 678 GZ2 morphological classes into 12 classes by excluding the detailed features, such as merging, disturbance, and irregularity (Table \ref{tab1}). 
\COMs{The EFIGI \citep{Baillard2011} samples contained the morphological types for 4,458 galaxies ($z \lesssim 0.05$). The EFIGI cataloged bright ($-23 < Mg < -13$) galaxies for reliable classification. A group of eleven experts followed RC3-like definitions (Third Reference Catalog of Bright Galaxies; \citealt{deVaucouleurs1963}) and Hubble classification (EFIGI Hubble sequence; see Table 1 of \citealt{Baillard2011}) to determine 18 morphological types.} 
\COMs{NA10 contained 14,034 galaxies in SDSS DR4 at $0.01 < z < 0.1$ with $g < 16$ mag. NA10 contained 14 morphological types that followed RC3-like classification (see Table 1 of \citealt{Nair2010}). }

\begin{table}[ht!]
\centering
\caption{Adopted morphological types for GZ2 in this study} 
\begin{tabular}[t]{ll}
\hline
Morphology &Description \\
\hline
Er& Elliptical, smooth and completely rounded\\ 
Ei& Elliptical, smooth and in between roundness\\
Ec& Elliptical, smooth and cigar shaped\\
Edge on& Edge on\\ 
Sa& Spiral, no bar, dominant bulge\\
Sb& Spiral, no bar, obvious bulge\\ 
Sc& Spiral, no bar, just noticeable bulge\\
Sd& Spiral, no bar, no bulge\\
SBa& Spiral, bar, dominant bulge\\
SBb& Spiral, bar, obvious bulge\\
SBc& Spiral, bar, just noticeable bulge\\
SBd& Spiral, bar, no bulge\\
\hline
\end{tabular}
\label{tab1}
\end{table}%

\begin{figure}[ht!]
\epsscale{1.15}
\plotone{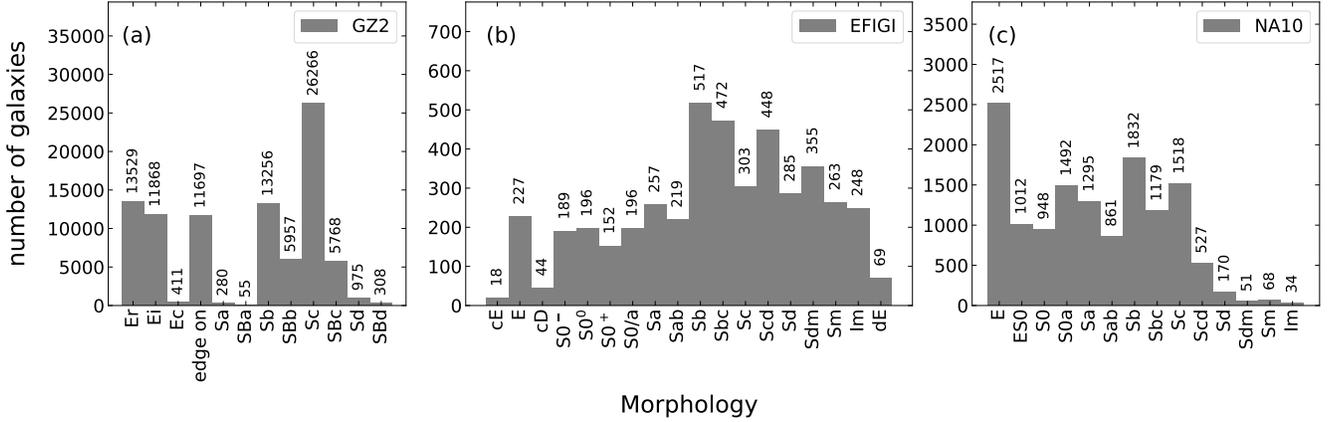}
\caption{Galaxy morphology distributions of GZ2 (a), EFIGI (b), and NA10 (c). The number of galaxies for each type is denoted at the top of each bar.}
\label{fig2}
\end{figure}

 \COM{Figure \ref{fig2} shows the morphological distribution of galaxies in the GZ2, NA10, and EFIGI catalogs. The GZ2 catalog has a significantly higher proportion of Sc-type spiral galaxies than Sa- and Sd-type galaxies. This is not the case for the NA10 and EFIGI catalogs. This suggests that the distribution of galaxy types in the GZ2 catalog may be affected by factors such as the expertise of the classifiers. For example, non-expert classifiers may be more likely to classify galaxies as Sc-type due to the compromise effect, which occurs when classifiers are unsure of the correct type and choose a more common type. This could lead to an underrepresentation of Sa- and Sd-type galaxies in the GZ2 catalog.}
\COMs{On the other hand, the EFIGI catalog exhibits a more balanced distribution of morphological types, except the relatively rare cE and cD types (Figure \ref{fig2}(b)). In contrast, the NA10 catalog has a notably lower proportion of irregular galaxy morphologies (Sd, Sdm, Sm, and Im) than the EFIGI catalog (Figure \ref{fig2}(c)).}

Identifying the detailed structural features (e.g., spiral arm, bar, and bulge) and accurate morphological classification from a low-spatial resolution galaxy image is challenging. \COM{Also, the absolute number and variety of images are very important to train the model.} Therefore, we employed the second phase of GZ2 \citep{Willett2013}, which listed high-spatial resolution galaxy images to determine the galaxy morphologies visually. 

\begin{figure}[ht!]
\epsscale{0.85}
\plotone{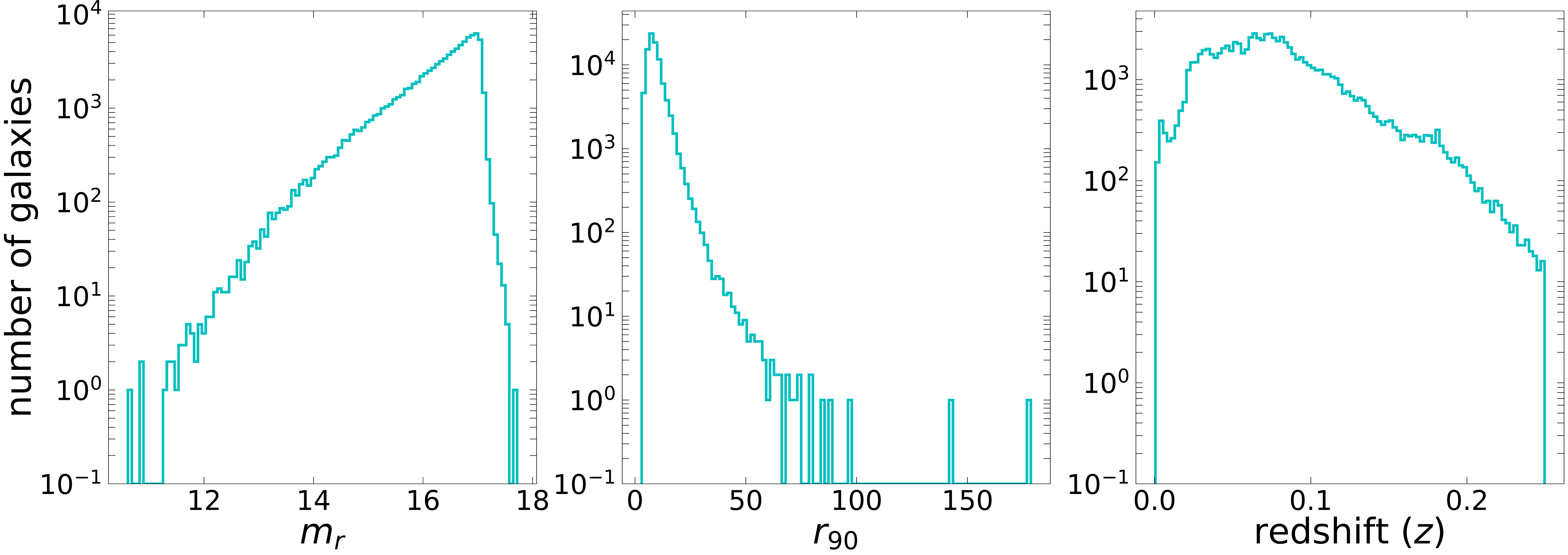}
\caption{Distribution of $m_r$ (left), $r_{90}$ (middle), and redshift (right) of the 90,370 selected galaxies from GZ2.
}
\label{fig1}
\end{figure}

Galaxy image samples listed in GZ2 were obtained from SDSS Data Release 7 \citep{Abazajian2009}, and the target galaxies were selected from the SDSS primary galaxy samples \citep{Strauss2002}. 
The selection criteria for galaxies in the GZ2 project are defined as follows: apparent magnitude ($m_{r}$) less than 17 mags (or less than 17.77 mags for Stripe82, \citealt{Willett2013}) and a Petrosian radius ($r_{90}$) greater than 3 $\arcsec$, with a redshift ($z$) between 0.0005 and 0.25 (Figure \ref{fig1}).
The total number of galaxies in GZ2 is 304,122. A total of 243,500 images exhibit spectroscopic redshift and reliable morphology. The images with 
flag\footnote[1]{The flag has a value of 1 when a) the votes and voter turnout for each question exceeded a set threshold and when b) the debiased vote fraction (corrected for classification bias) was 0.8 or higher. It was zero otherwise. Therefore, if the flag were 1, it would be closer to a cleaner sample.}\COM{$=$1} and classified as either E (smooth) or S (features or disk) galaxies were selected. The galaxy images contaminated by bright stars and unreliable background measurements were excluded. A total of 90,370 (25,808 and 64,562, for E and S galaxies, respectively) galaxy images were finally selected for ML. 

\subsection{Data Preprocessing} \label{sec:dataPrep}

We downloaded the $gri$ galaxy images in \COM{FITS} format from SDSS. The images were photometrically preprocessed, and the background was subtracted. 
We applied a preprocessing of the images for ML as follows.

\begin{figure}[ht!]
\epsscale{1.15}
\plotone{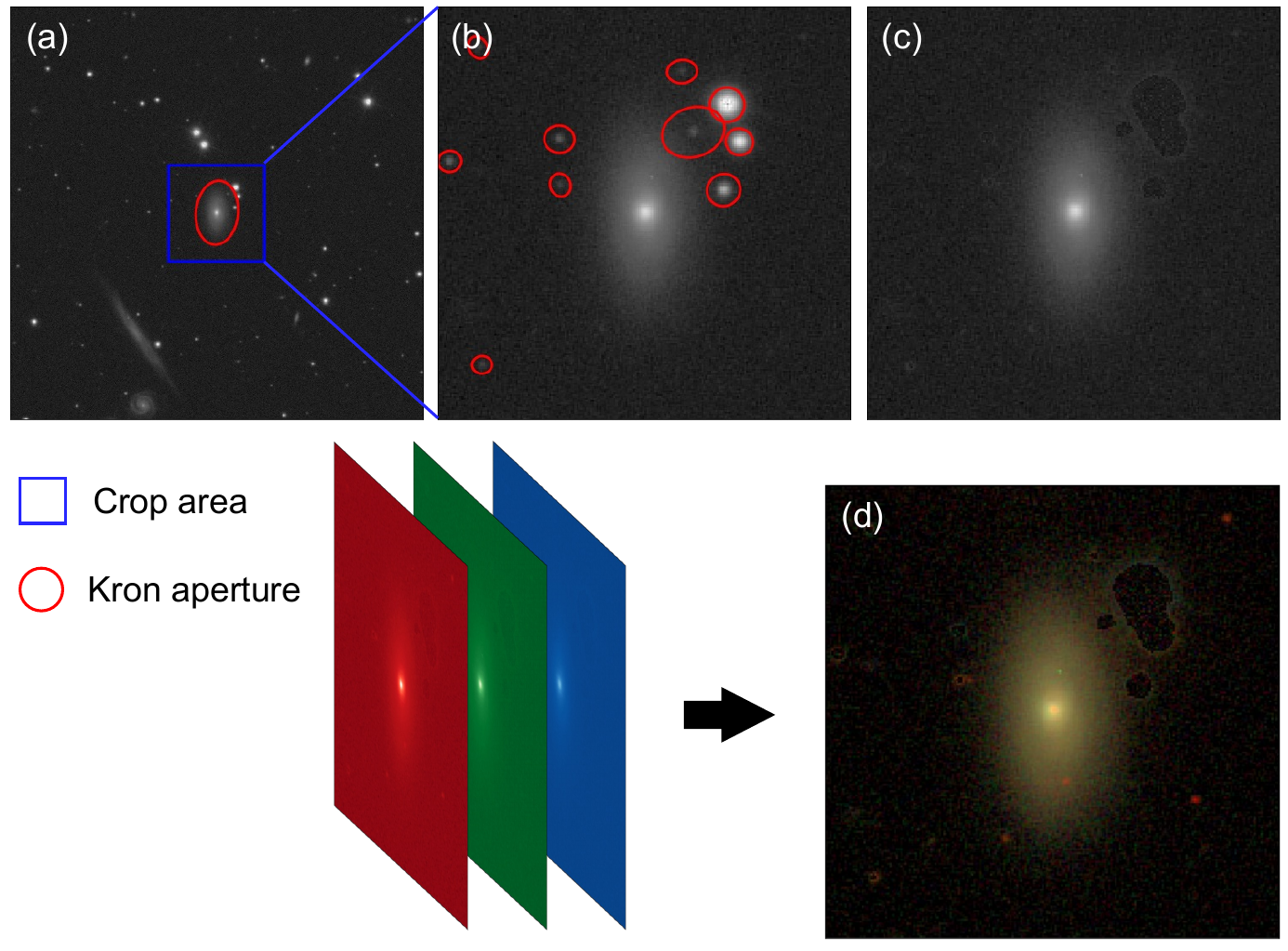}
\caption{Data preprocessing method. (a) The downloaded image depicts the target galaxy in the red ellipse. The blue square depicts three times the kron radius along the semi-major axis. (b) The cropped image depicts the non-target objects in the red ellipses. (c) The cropped image depicts the masked non-target objects. (d) RGB color image.    
}
\label{fig3}
\end{figure}

The target galaxy was located in the downloaded image’s center. The image was cropped up to three times the target galaxy’s \COM{Kron} radius (Figure \ref{fig3} (a), (b)). The kron radius was calculated from the $r$-band image using the Astropy \COM{photutils package \citep{Bradley2016}.} The galaxy’s apparent size ratio to the input image size was kept constant. All the other non-target objects identified by the Python \COM{photutils} package in the image were masked by filling randomly selected background values (Figure \ref{fig3} (b), (c)). Since ML requires fixed-sized images, the images were converted to 256×256 pixel size using the \COM{PyTorch Torchvision library \citep{Paszke2019}.}

With the cropped and masked $gri$ images, RGB color images were created using make\_lupton\_rgb, a subpackage of astropy.visualization. The color images were rendered by 0.5 linear stretch and 10 asinh softening to match the SDSS color images. The $g$-, $r$-, and $i$-band fluxes were multiplied by 1.1, 1, and 0.79, respectively (Figure \ref{fig3}(d)) to convert them into RGB color images. The pixel values were normalized between 0 and 255. 

\section{METHODS} \label{sec:method}
\subsection{Autoencoder Architecture}

\begin{figure}[ht!]
\plotone{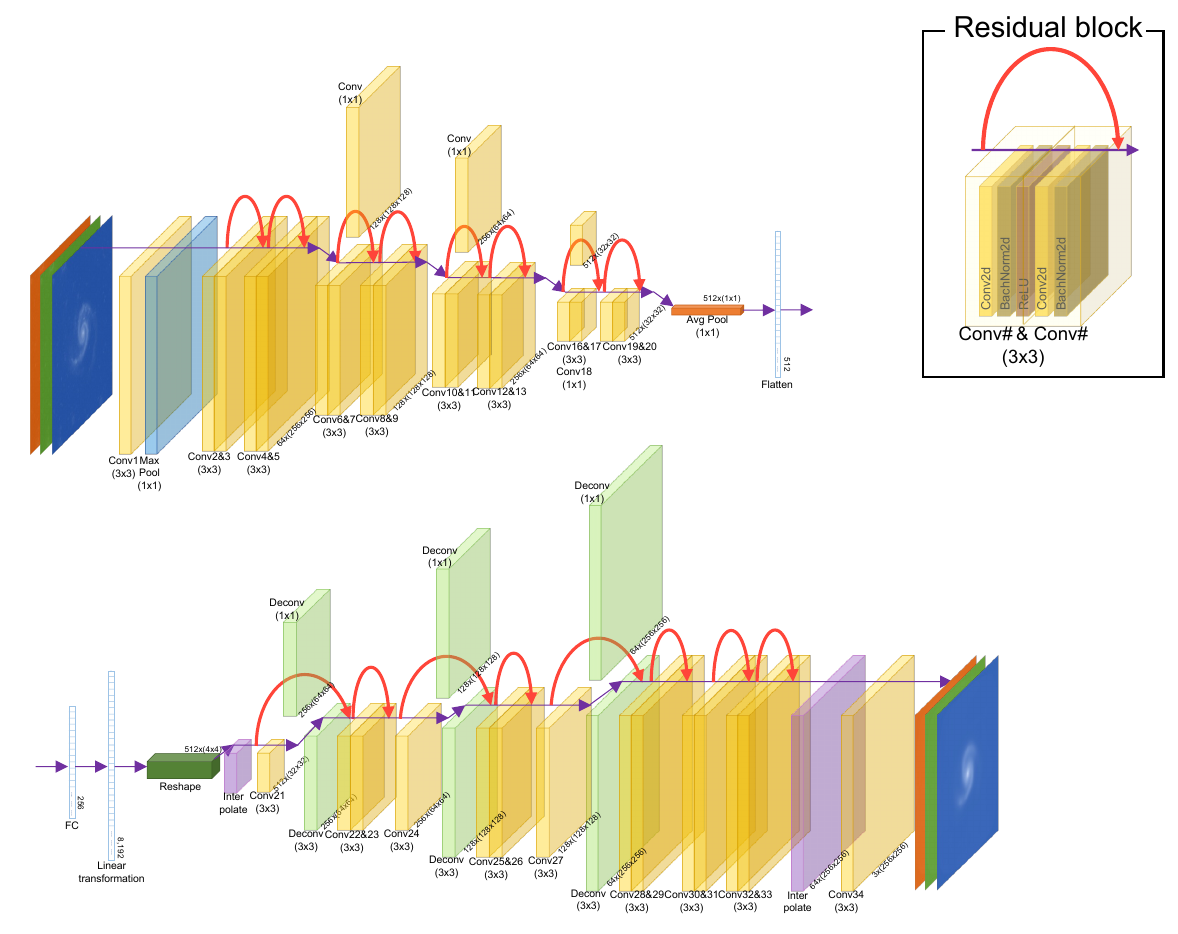}
\caption{Schematic of our convolutional autoencoder adopting ResNet-18. The sizes of the image are denoted below each layer.  
}
\label{fig4}
\end{figure}

We utilized the CAE, composed of an encoder and decoder, to minimize the difference between input and output images through the optimization of neural weights and biases in the fully-connected and convolutional layers.
The encoder adopts the residual neural networks-18 \citep[ResNet-18;][]{He2015}, which is composed of a convolutional layer, eight residual blocks, and a fully-connected layer (Figure \ref{fig4}). The residual block consists of two convolutional layers, followed by a batch normalization and rectified linear unit (ReLU) activation function for each convolutional layer \citep{Zeiler2013}. The gradient vanishing problem is reduced by residual learning, using a skip connection for each residual block \citep{He2015}. As the 256 × 256 × 3 input images were sequentially passed through the convolution layers and residual blocks, they were transformed into 32 × 32 × 512 images. A total of 512 latent features were extracted via adaptive average pooling. 

The decoder architecture was an apparent reversal of the encoder’s architecture. The decoder was composed of one convolution layer and eight residual blocks. The encoder’s 512 latent features output as input to the decoder. The decoder’s input images were converted into 256 vectors as the fully-connected layers. Linear transformation and reshaping resulted in 4 × 4 × 512 images, which were subsequently expanded by interpolation to 32 × 32 × 512. The images were passed through the residual block composed of the convolutional and de-convolutional layers and a three-convolution layer block, followed by the seventh and eighth blocks composed only of the convolutional layers. The images were passed through a convolution layer to convert them into the inputted image dimensions, 256 × 256 × 3.  

\subsection{Training and Validation}

 The source code was written in Python. Pytorch and Pytorch Lightning \citep{Falcon2020} were used for the back-ends. The training loop was processed in Pytorch Lightning. Experiments and training were performed on four multi–NVIDIA TITAN RTX D6 graphic processing units (GPUs) with 24 GB RAM and CUDA-11.4. A total of 90,370 images were divided into training, validation, and test datasets in the 3:1:1 ratio. We adopted batch size (42), learning rate (0.0001), loss function (mean squared error; MSE), and optimizer (Adaptive Moment Estimation; Adam) parameters for the training and validation. 
  For additional options, we used imbalanced dataset sampler\footnote[2]{\url{https://github.com/ufoym/imbalanced-dataset-sampler}} to contain balanced morphological types in each batch. Furthermore, the images are randomly rotated\footnote[3]{torchivision.transforms.RandomRotation is used.} before being used as input for each training iteration to reduce dependence on position angle.
  


\begin{figure}[ht!]
\epsscale{1.0}
\plotone{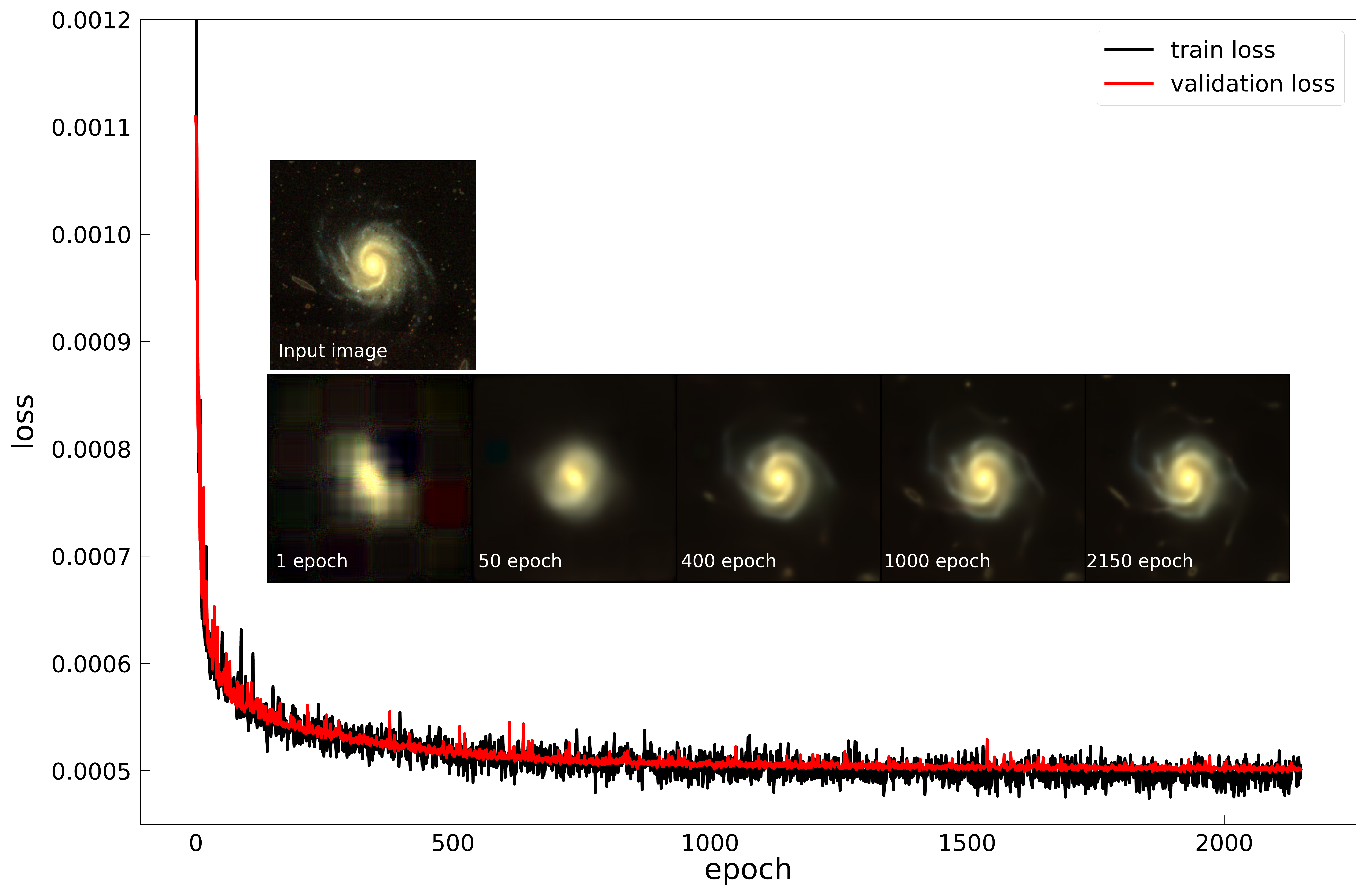}
\caption{Training and validation losses as a function of epoch in the CAE. The insets depict the input and output images from the CAE at indicated epochs.   
}
\label{fig5}
\end{figure}

Figure \ref{fig5} depicts the training and validation loss, which decreased during the training progresses but did not decrease dramatically after a thousand epochs. Thus, we stopped the training at the 2150$^{th}$ epoch and adopted the final training as the final CAE model. We compare the input and the output images at a given epoch in inset images. 

\begin{figure}[ht!]
\epsscale{1.1}
\plotone{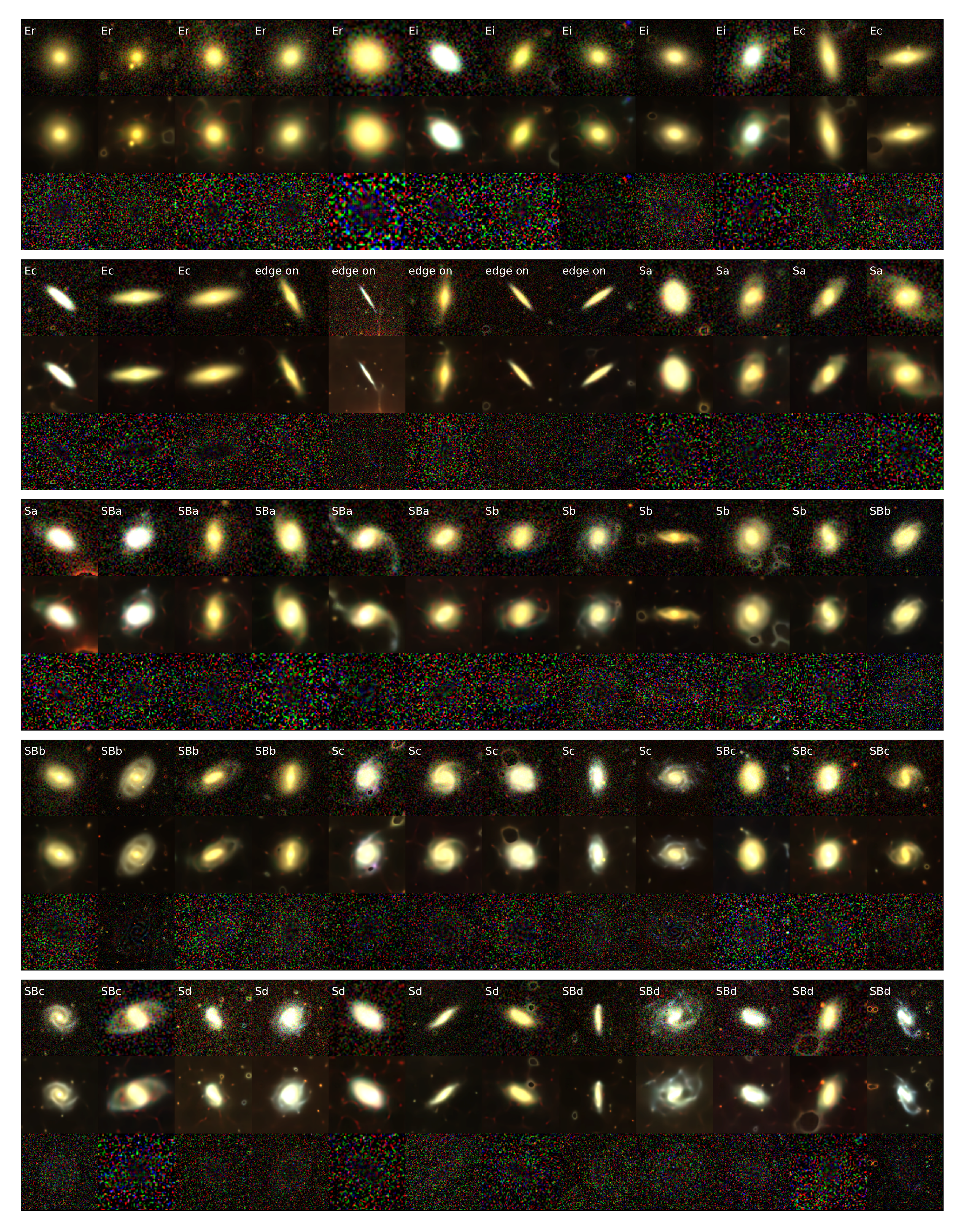}
\caption{Top, middle, and bottom line depict the test sample images, output image reconstructed by the CAE, and the residuals, respectively.      
}
\label{fig6}
\end{figure}

Figure \ref{fig6} depicts the input image (test dataset; top line of each panel), the output image (middle line of each panel), and the residual image (bottom line of each panel) for evaluating the accuracy of the final CAE. The input and output images appeared similar, although the output images were slightly smoother. None of the galaxy shapes or features remained in the residual images, regardless of their morphological types. The median and standard deviation of the mean pixel values for the residual images of all test datasets were 0.0181 and 0.003, respectively. It means that the model's encoder effectively captures the essential features of the input images.

\subsection{Database of Latent Features}

\begin{figure}[ht!]
\epsscale{0.85}
\plotone{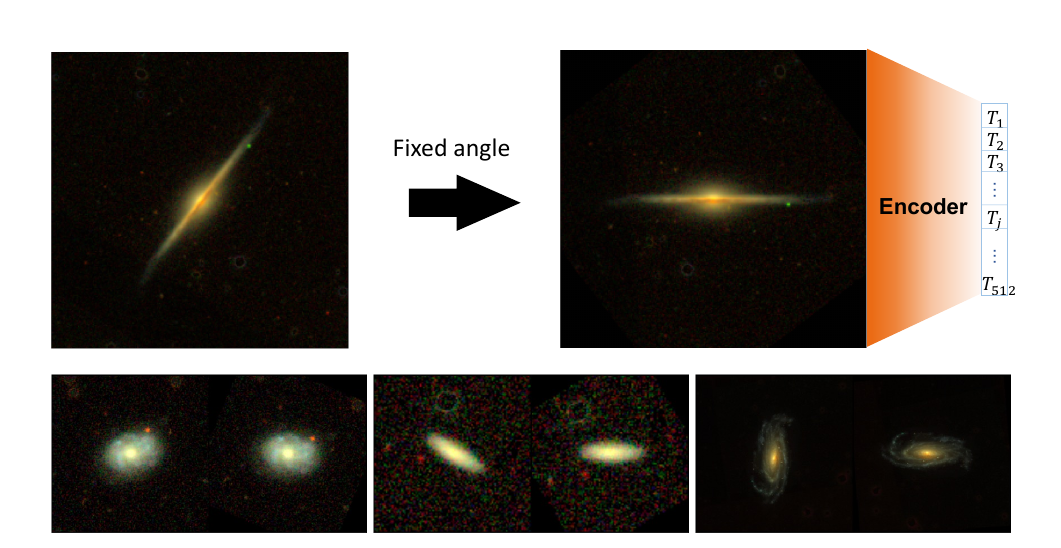}
\caption{Example of rotating the position angle of 90° before constructing DB.}
\label{fig7}
\end{figure}

The final CAE successfully reproduced the output galaxy images from the input galaxy images, suggesting that the 512 latent features extracted by the encoder output contain galaxy morphological features. Before creating the latent features, we must make a position angle of 90 degrees for all input galaxy images (see Figure \ref{fig7}) to prevent the position angle dependency on searching similar galaxy images 
 \COM{where the original position angle was determined by SouceCatalog class of Photoutils package.}
The objective of randomly rotating images during training differs from maintaining a fixed position angle for similarity in morphological classification. The model is intended to be insensitive to the galaxy's orientation, but to identify similar images, the position angle must be controlled. The encoder created the latent feature databases from GZ2 (90,370), EFIGI (4458), and NA10 (14,034) input galaxy images. 

\subsection{Similar Image Retrieval}

The traditional morphological classification relies on limited non-parametric features, such as concentration, asymmetry, density, Gini coefficient, and M20, to represent galaxy structures in input images. However, these few features may not fully capture the complexity of galaxy images. A more comprehensive set of galaxy image parameters is needed to identify similar galaxy images accurately. Therefore, we employed 512 latent features as the galaxy morphological parameters. 

\begin{figure}[ht!]
\epsscale{1.15}
\plotone{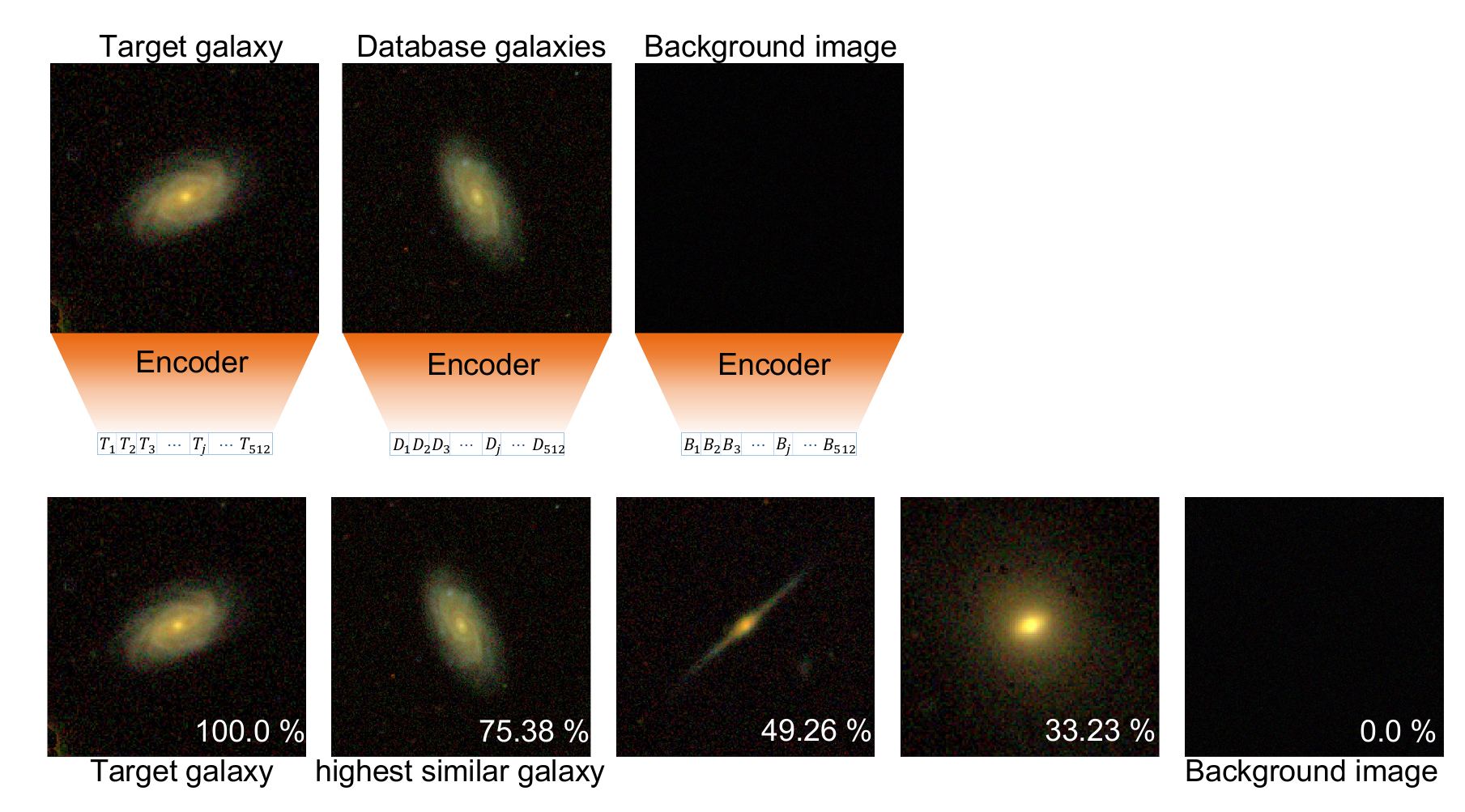}
\caption{Schematic showing the similarity parameter calculation. The top and bottom panels display the encoder output and the galaxies with different similarity levels, respectively.}
\label{fig8}
\end{figure}

With a given target galaxy, a similarity parameter ($Sp$) was defined to retrieve a similar galaxy from the database (Figure \ref{fig8}). 

\begin{equation}
Sp_i = \left( 1 - \frac{EN_{gal,i}}{EN_{back}} \right) \times 100
\end{equation}

\begin{equation}
EN_{gal,i} = \sqrt{\sum_{j=1}^{512} \ (T_j - D_{ij})^2}
\end{equation}

\begin{equation}
EN_{back} = \sqrt{\sum_{j=1}^{512} \ (T_j - B_j)^2}
\end{equation}

$T_j$, $D_{ij}$, and $B_j$ are the j'th latent feature of the target galaxy, of the i'th database-listed galaxy, and of the background of the target galaxy image, respectively. 
We designate the Euclidean norm (EN) for each galaxy listed in the database as $EN_{gal,i}$, and for the background image of the target galaxy as $EN_{back}$.
The background images were created by randomly selecting the region outside the objects in the target galaxy image.  

A similar galaxy was determined as a database galaxy with the highest similarity parameter (H-similarity). It should be noted that if the similarity is 100\% (meaning the exact same galaxy image as the target image in the database), we choose the galaxy with the second-highest similarity parameter. 

\begin{figure}[ht!]
\epsscale{0.4}
\plotone{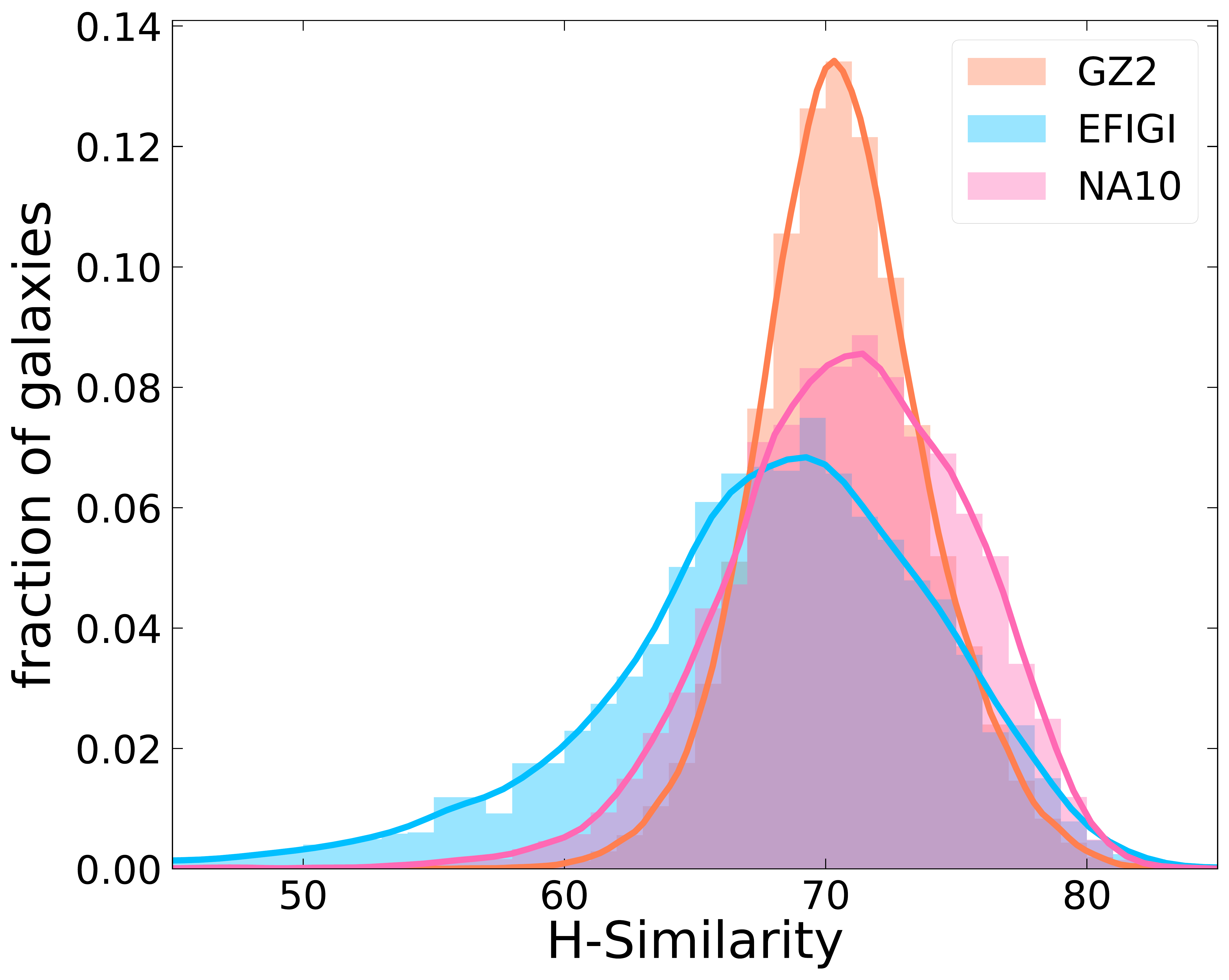}
\caption{Distribution of the H-similarity of each morphology catalog.}
\label{fig9}
\end{figure}

The H-similarity is calculated for all galaxies in each catalog for a given target galaxy.  Therefore, the number of the H-similarity values calculated for a target galaxy equals the number of galaxies in each catalog.
Figure \ref{fig9} depicts the distribution of H-similarity for each catalog. 
In the comparison of the three catalogs, both GZ2 (70.6$\%$ $\pm$ 3.25$\%$) and NA10 (70.8$\%$ $\pm$ 4.62$\%$) show slightly higher average H-similarity scores than EFIGI (67.7$\%$ $\pm$ 6.88$\%$). \COM{Notably, the H-similarity score distribution in the EFIGI catalog is skewed towards lower scores. This may be due to the smaller size of the EFIGI catalog, which decreases the probability of finding high-similarity pairs, and the lower fraction of early-type galaxies in the EFIGI catalog, which tend to have higher H-similarity scores (see section \ref{sec:H-similarty_morphology}).}
Further, if the H-similarity of the target galaxy is less than one sigma from the average of the H-similarity in each catalog, it is decided that there is no similar image in the database. It should be noted that some galaxies with low H-similarity might have complex structures that make it difficult to find similar galaxies in the database.  

\section{Morphological Classification} \label{sec:classify}

After training CAE and finding a similar galaxy in the database, we classified the morphology of the target galaxy, adopting the similar galaxy's morphological type. The morphological types in the database have come from GZ2, EFIGI, and NA10 catalogs.

\subsection{H-similarity by Morphology} \label{sec:H-similarty_morphology}

\begin{figure}[ht!]
\epsscale{1.15}
\plotone{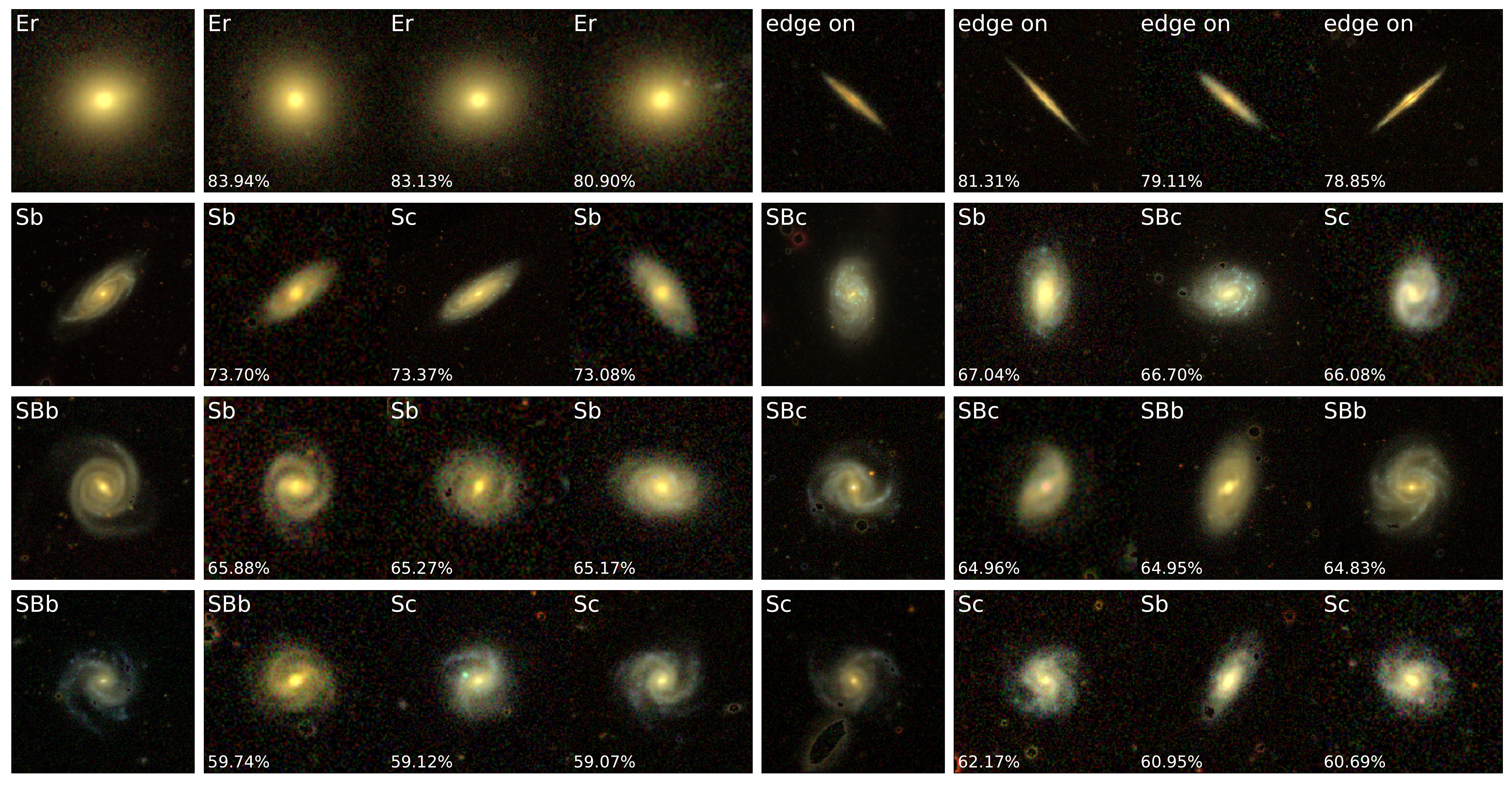}
\caption{Examples of \sk{the retrieved galaxies with morphological types and similarity.} The leftmost image depicts the target galaxy in each morphology. Subsequent images from the left depict the top three highly similar galaxies. Letters and percentages indicate the morphological type and similarity, respectively.
}
\label{fig10}
\end{figure}

\begin{figure}[ht!]
\epsscale{0.8}
\plotone{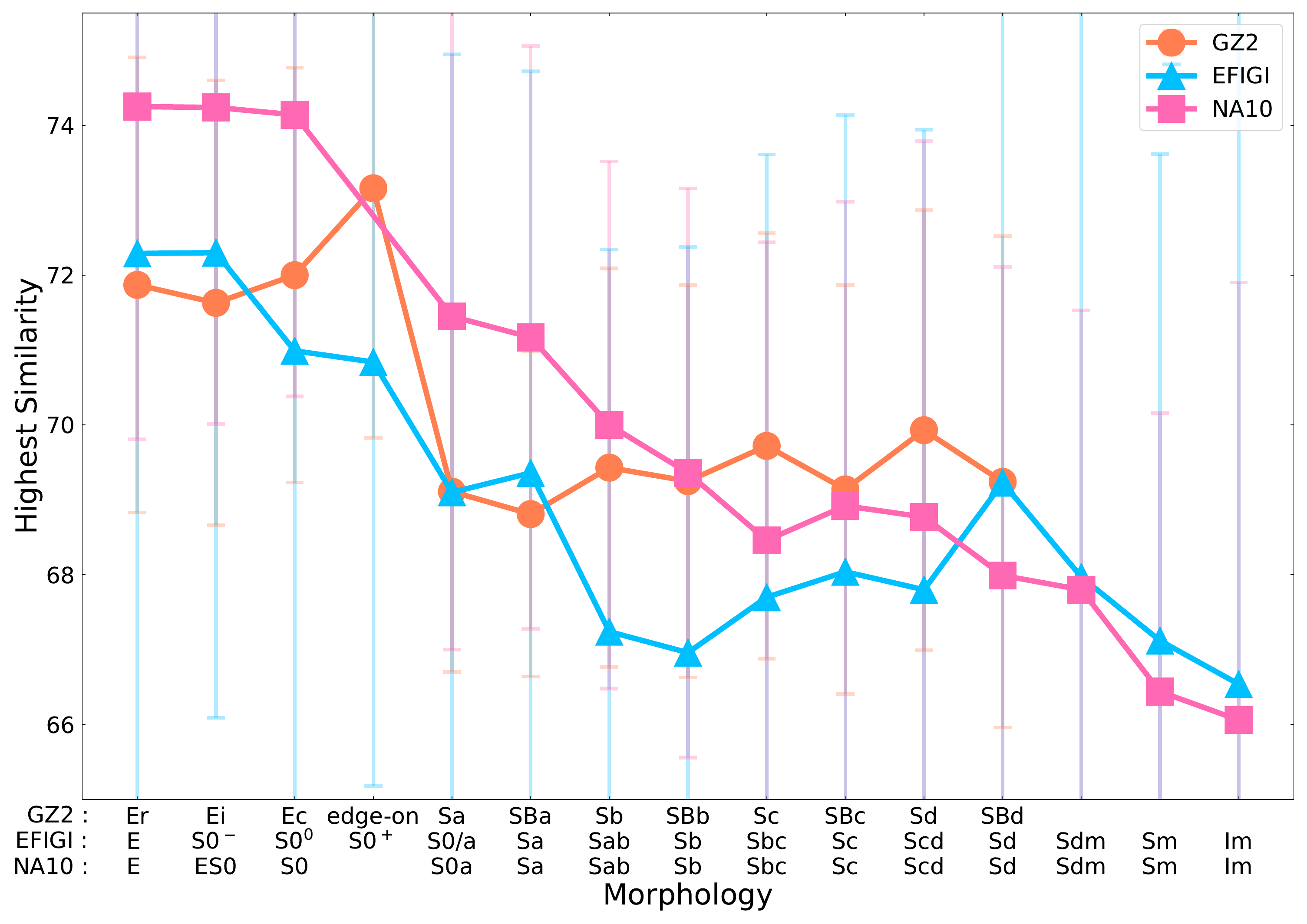}
\caption{H-similarity as a function of the morphological types in GZ2 (orange circles), EFIGI (cyan triangles), and NA10 (pink squares) catalogs. The error bars represent the standard deviations.}
\label{fig11}
\end{figure}

Figure \ref{fig10} depicts the galaxy morphologies of the retrieved galaxies and similarity indices. Simple-structure galaxies, such as elliptical or edge-on galaxies, exhibited larger H-similarity values. The H-similarity decreased as the structural complexity of galaxies increased, such as face-on late-type galaxies. Figure \ref{fig11} shows the H-similarity as a function of the galaxy morphology for each catalog. The average of the H-similarity value decreases from early- to late-type as the internal structure
becomes complex, suggesting that the structural features (e.g., bulge, arms, bars, rings) and the constituent properties (e.g., dust, star-forming region) of the early- and late-type galaxies affect their similarity.

\subsection{Predicted Morphology}
\subsubsection{Morphological classification accuracy}

To evaluate the reliability of morphological classification using TSGICAS, we divided the database into the morphology-check (MC) sample (10$\%$) and morphology-Database (MD) sample (90$\%$) in each morphological type. We calculated the H-similarity for each MC sample within the MD samples and determined the MC sample morphology. 

\begin{figure}[ht!]
\epsscale{1.15}
\plotone{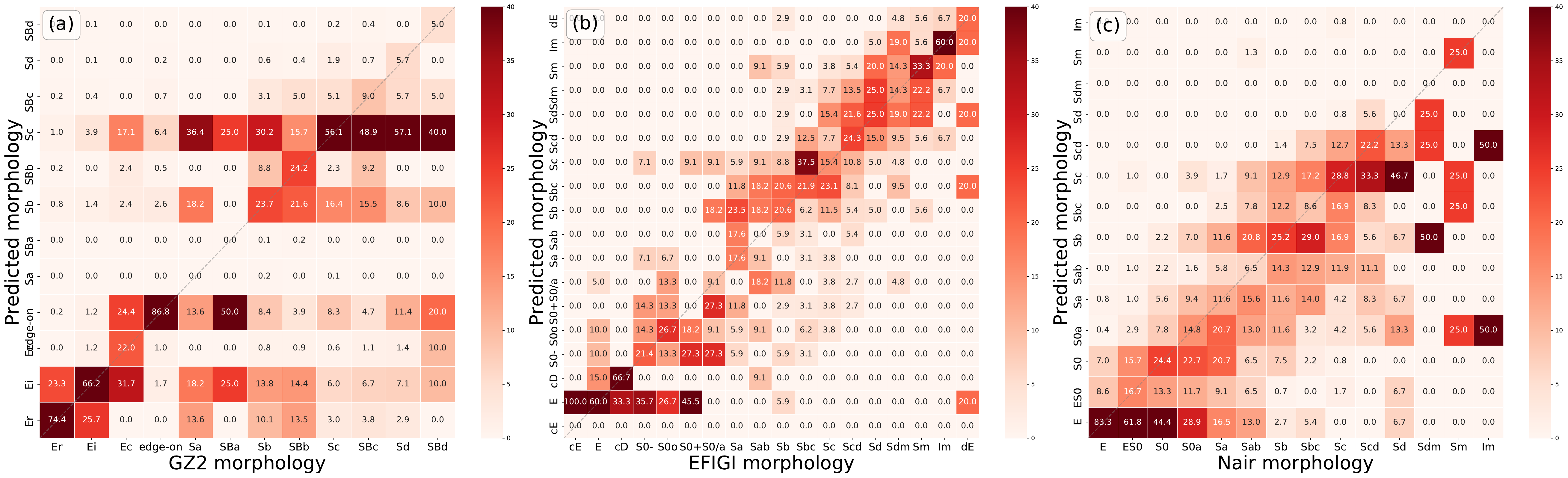}
\caption{Confusion matrices compare identified and predicted morphological types in the catalogs. The columns represent the instances of each morphological type in the catalogs, while the rows represent the instances of the predicted morphological types. The values within each matrix block represent the percentage of individuals for a given identified morphological type. The dashed line represents the perfect match between the two morphological types.}
\label{fig12}
\end{figure}

Figure \ref{fig12} depicts the MC sample morphological confusion matrices for the three morphological catalogs. Our morphological classification generally predicts a reliable morphological type for the MC sample. Most of the predicted morphological and morphological types of MC samples in confusion matrices show a one-to-one correlation, although it shows some scatters. Correlation coefficients are 0.687, 0.819, and 0.784 in GZ2, EFIGI, and NA10, respectively. 

\begin{figure}[ht!]
\epsscale{1.15}
\plotone{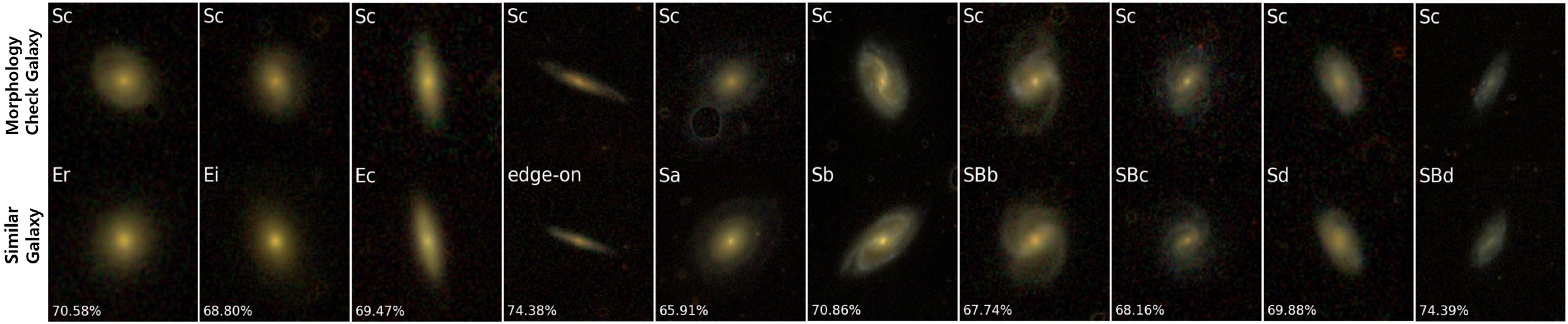}
\caption{Examples of Sc-type galaxy miss-classification in GZ2 catalog. The top and bottom rows indicate the MC sample’s Sc type galaxies and the retrieved similar images. The text and percentage denote the morphology type and similarity values, respectively.}
\label{fig13}
\end{figure}

In the three catalogs, GZ2 samples exhibit significantly large scatters at the spiral galaxies and are biased to Sc types (Figure \ref{fig12}(a)). EFIGI (Figure \ref{fig12}(b)) and NA10 (Figure \ref{fig12}(c)) sample classification are comparable to the reported visual morphological classification (see Figure 19 in \citealt{Baillard2011} and Figure 14 in \citealt{Nair2010}). This biased classification of GZ2 through our TSGICAS seems to result from the difficulty of visual morphological classification among Sa to Sd types by non-experts. In Figure \ref{fig2}, GZ2 catalog hosted a higher number of Sc galaxies than the adjacent spiral galaxies, while EFIGI and NA10 catalogs hosted comparable numbers. Figure \ref{fig13} depicts an example in which similar images of Sc type have different types in GZ2. The two galaxies are indistinguishable by visual inspection, indicating that Sc type contains too many different types of galaxies. Therefore, we suspect that morphological types of Sc galaxies in GZ2 are confused mainly with the adjacent types of spiral galaxies.  

To investigate this issue, we randomly select a small and fixed number of samples in each morphological type to make a new MD sample. 
Assuming the original MD sample has a well-normalized distribution of classification accuracy, it is expected that the new MD sample will exhibit a lower variance in accuracy, indicating a higher probability of well-classified galaxy types in the new MD sample compared to the original sample.
We calculated the H-similarity and determined the morphological types for the MC samples in the new MD samples. This process was iterated 400 times. For each galaxy in the MC sample, we finally determine a morphological type as a mode value of 400 predicted morphological types. 
This will statistically reduce the probability of being misclassified by repeatedly performing galaxy classification in the new MD with a low variance of classification accuracy.

\begin{figure}[ht!]
\epsscale{1.15}
\plotone{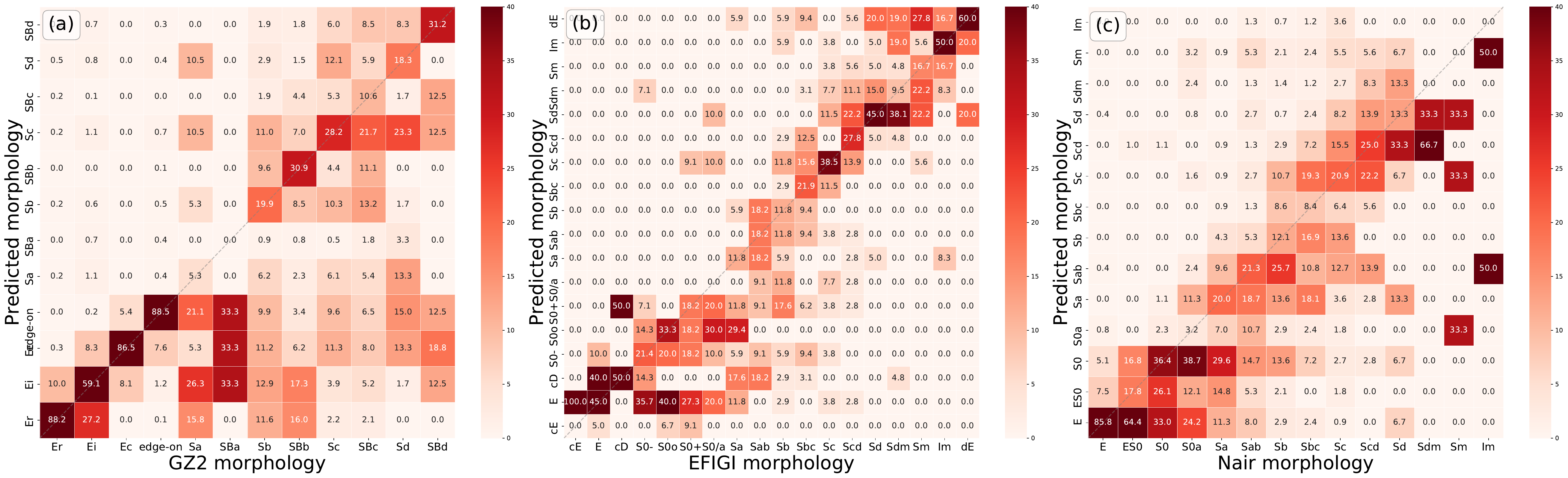}
\caption{Confusion matrices compare morphological types for each catalog, along with the mode value of the predicted morphological type for 400 newly generated MD samples. The values within each matrix block represent the percentage of individuals for a given identified morphological type. The dashed line represents the perfect match between the two morphological types.}

\label{fig14}
\end{figure}

\COM{Figure \ref{fig14} presents the confusion matrices of this test. The bias of the predicted morphological type for spiral galaxies in the GZ2 catalog is lower than that of the previous confusion matrices (Figure \ref{fig12}). The correlation coefficients for GZ2 (0.735), EFIGI (0.811), and NA10 (0.815) catalogs are either higher or comparable with previous studies. The difficulty of morphological classification among adjacent morphological types via visual classification translates into the uncertainty of predicted morphological types by TSGICAS. As a result, GZ2 and NA10 exhibited improved results in late-type galaxies, indicating the presence of a significant number of misclassified galaxies within the spiral region. In conclusion, a well-classified morphological catalog is essential for more reliable predictions by TSGICAS.}


\subsubsection{Misclassification}\label{Misclass}

To identify galaxies that were misclassified by our previous morphological classification method, we have divided the results into three categories based on the degree of misclassification: \emph{best-classification}, \emph{good-classification}, and \emph{miss-classification}. 
\emph{Best-classification} refers to instances where the morphological type assigned to the MC sample by our method exactly matches the true morphological type. 
\emph{Good-classification} refers to instances where the predicted morphological type is within two adjacent types of the true type and encompasses the category of \emph{best-classification}.
For instance, \emph{good-classification} of Sc type in the EFIGI catalog include Sb, Sbc, Sc, Scd, and Sd.
\emph{Miss-classification} refers to all remaining cases that are not included in the \emph{best-classification} and \emph{good-classification}.
Our method provided the \emph{best-classification} of 55.08$\%$ (GZ2), 24.44$\%$ (EFIGI), and 31.74$\%$ (NA10) and the prediction accuracies of the \emph{good-classification} are 82.68$\%$ (GZ2), 74.28$\%$ (EFIGI), and 80.32$\%$ (NA10). The fractions of the \emph{miss-classification} are 17.32$\%$ (GZ2), 25.72$\%$ (EFIGI), and 19.68$\%$ (NA10).



\begin{figure}[ht!]
\epsscale{1.17}
\plotone{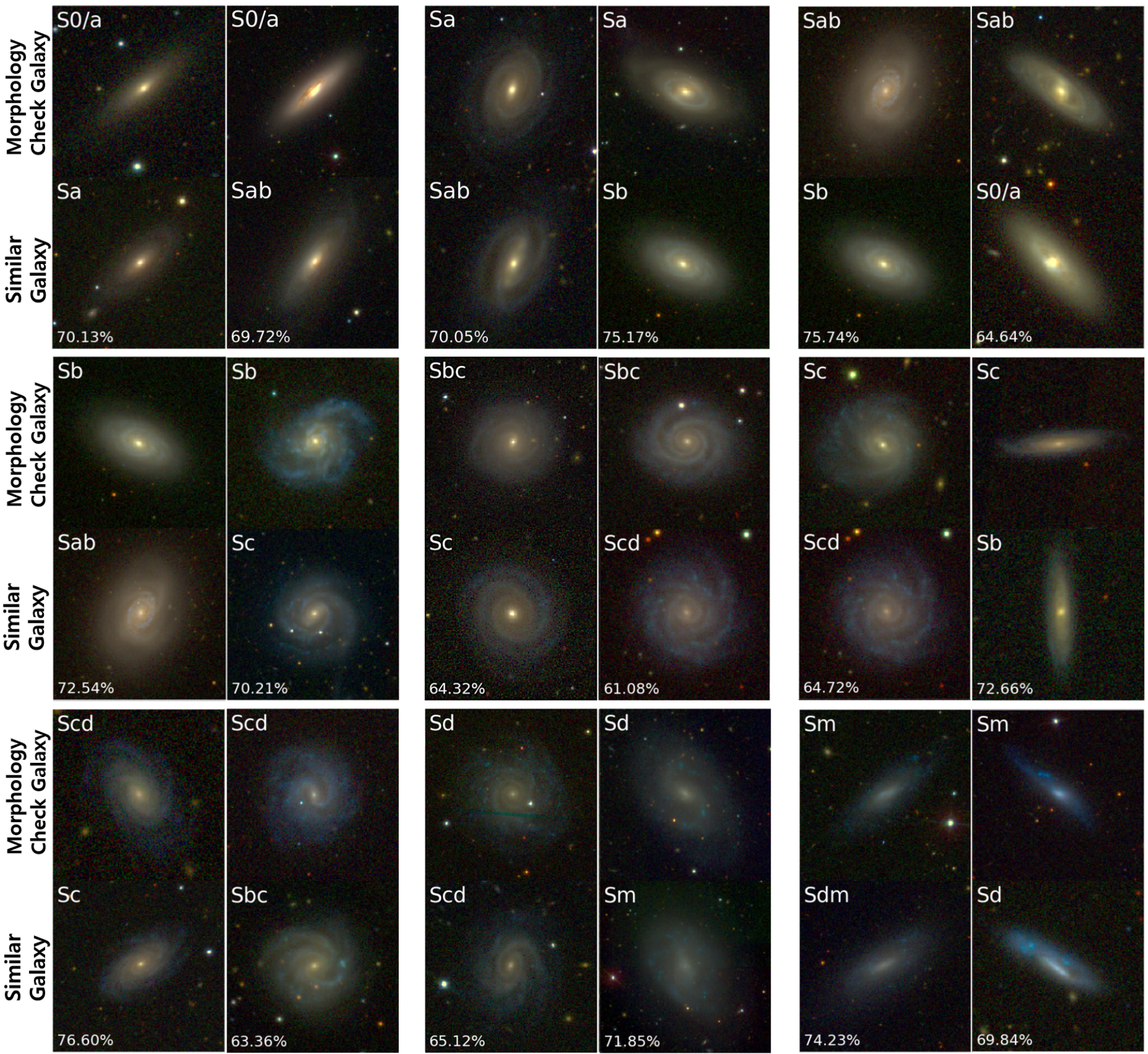}
\caption{Examples of \textbf{\emph{good-classification}} among the late-type galaxies. The top and bottom galaxy images in each figure represent the MC samples and retrieved similar images, respectively. Each set's first and second columns depict the 1$^{st}$ and 2$^{nd}$ adjacent morphological types, respectively. The text and percentage indicate the morphological types and similarity values, respectively.  
}
\label{fig15}
\end{figure}

\begin{figure}[ht!]
\epsscale{1.15}
\plotone{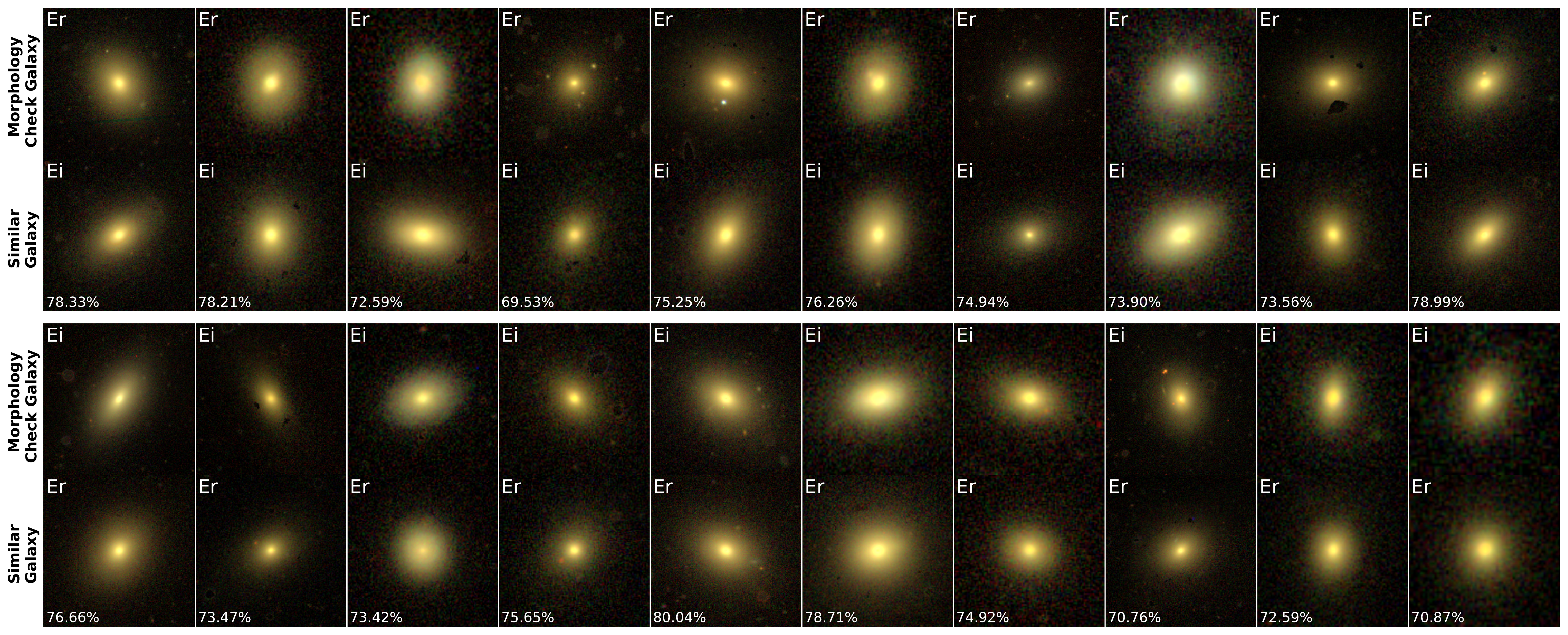}
\caption{Examples of \emph{good-classification} of the early-type galaxies in GZ2 catalogue. The top and bottom images display the MC samples and the retrieved similar images, respectively. The text and percentage indicate the morphological types and similarity values, respectively. 
}
\label{fig16}
\end{figure}

\begin{figure}[ht!]
\epsscale{1.15}
\plotone{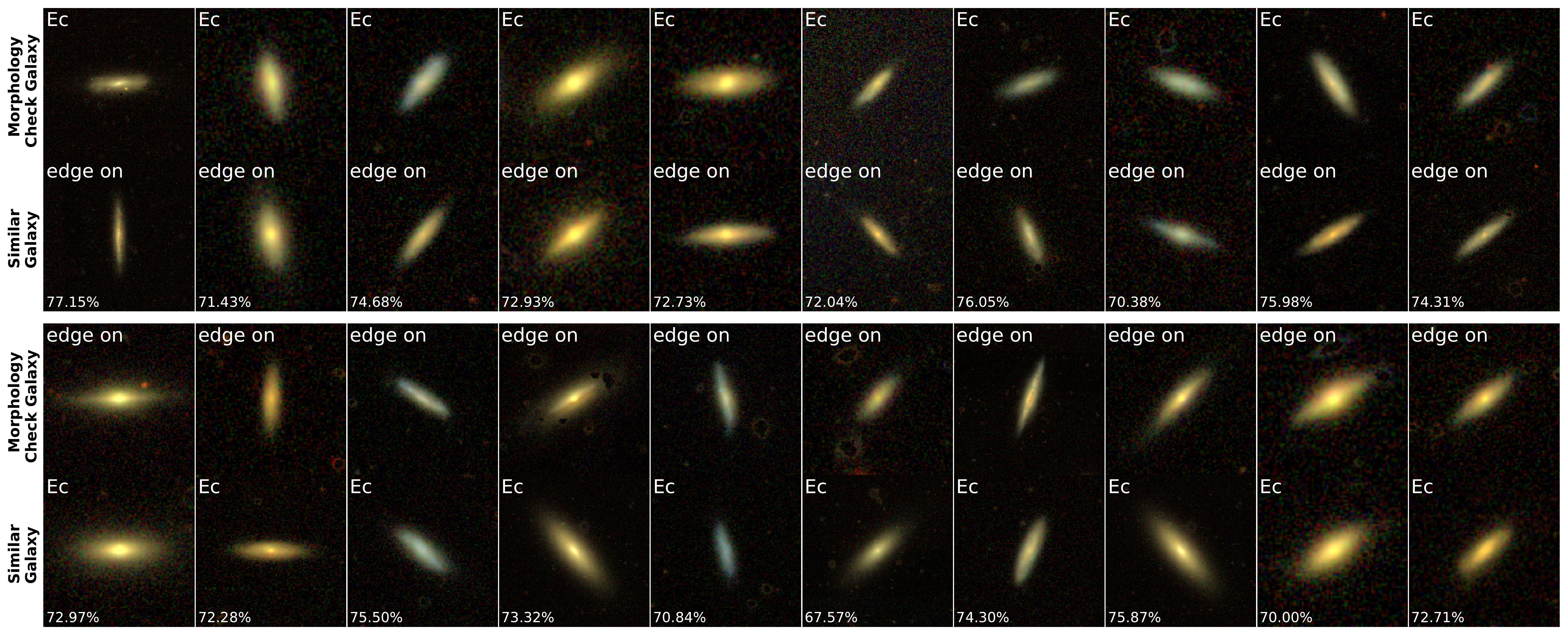}
\caption{Example of galaxy in the \emph{good-classification} category that lies on the boundary between early-type (Ec) and late-type (edge-on) galaxies in the GZ2 catalog. The top and bottom images display the MC samples and the retrieved similar images, respectively. The text and percentage indicate the morphological types and similarity values, respectively.
}
\label{fig17}
\end{figure}

Figures \ref{fig15} and \ref{fig16} depict examples of \emph{good-classification} galaxies for late-type and early-type galaxies, respectively. 
Furthermore, Figure \ref{fig17} highlights that, despite being part of our definition of \emph{good-classification}, early-type (Ec) galaxies were misclassified as edge-on (late-type) galaxies and vice versa.
The examples in the \emph{good-classification} category demonstrate that distinguishing between adjacent morphological types can be challenging even for human observers, as the MC sample and similar galaxies have different morphological types but visually similar images.

\begin{figure}[ht!]
\epsscale{1.15}
\plotone{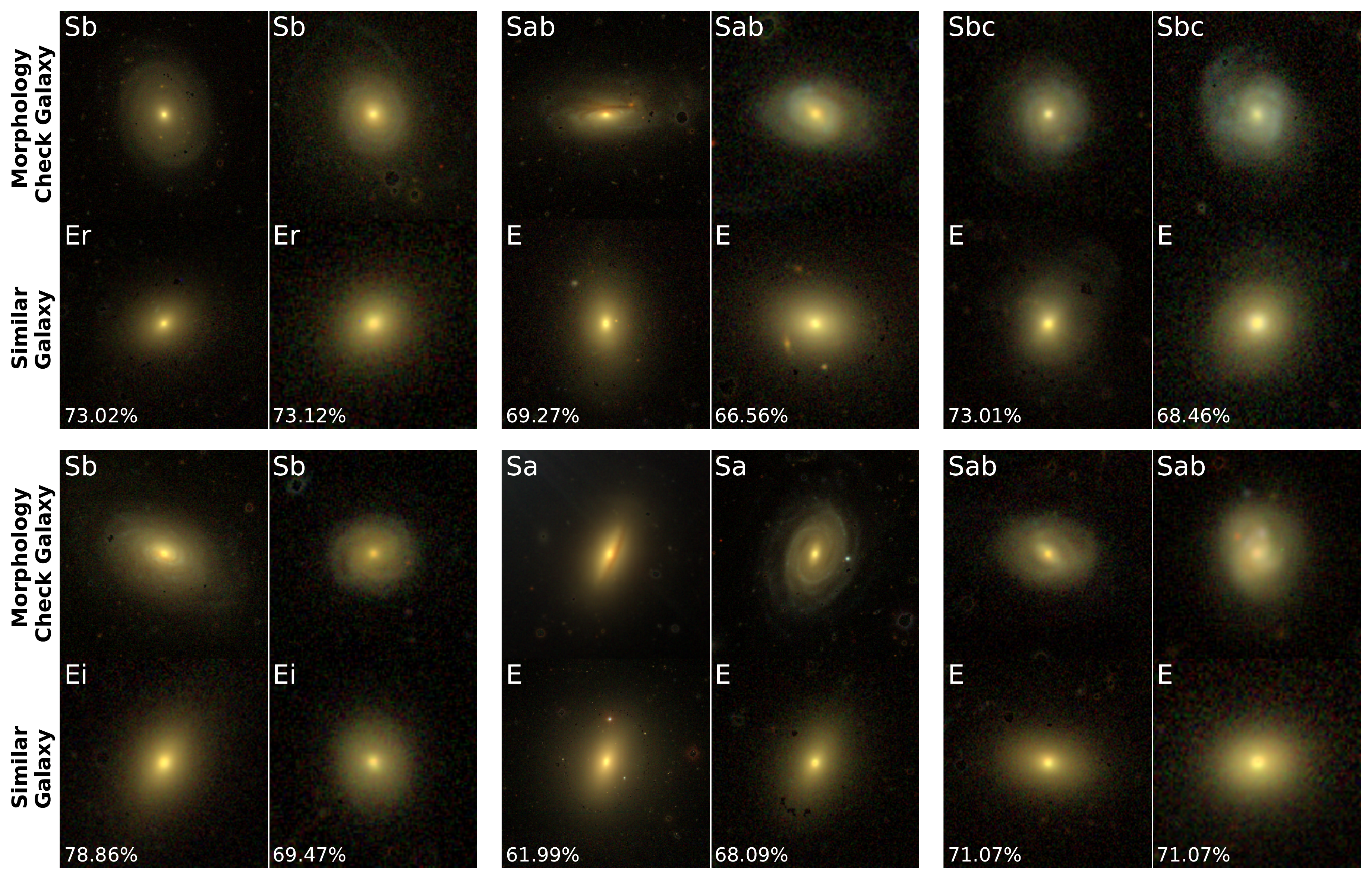}
\caption{Examples of \emph{miss-classifications} of late-type galaxies as early-type in each catalog. The top and bottom rows depict the MC samples and the retrieved similar images, respectively. The text and percentage indicate the morphological types and similarity values, respectively.
}
\label{fig18}
\end{figure}

Examples of the \emph{miss-classifications} are presented in Figure \ref{fig18}. Although these galaxies are late-type, the overall color was red similar to early-type, and their spiral arm structures were fainter than that of typical spiral galaxies. Moreover, galaxies with a high degree of similarity exist even though they do not appear structurally similar. This was our encoder’s limit. In these cases, the latent features appear to be biased toward photometric parameters representing the global shape or color of the galaxy rather than other parameters representing the detailed structures (etc., spiral arms, bared, rings) of the galaxy. For example, the TSGICAS prefers to determine red elliptical galaxies as similar images on red spiral galaxies with faint spiral arms since the latent features are biased to color rather than the spiral arm structure. The training image should be changed to a galaxy image with clear structures to solve this problem.

\section{Directions for Improvement}\label{sec:improve}

\COM{To refine our methodology, two key factors need to be addressed. First, it is crucial to provide a dataset of latent features that is balanced in terms of the accurate classification of galaxy morphologies. If the dataset size or its diversity is not enough, the chance of finding similar images decreases, thereby reducing the similarity of the retrieved images. Consequently, for TSGICAS to accurately classify galaxy morphology, it necessitates a balanced dataset where each morphological type of galaxy is equally represented. This is especially pertinent for galaxies with more complex structures, which require a diverse range of image data. To this end, we aim to increase the number of accurately classified galaxies in morphological categories that are currently underrepresented in the dataset, to at least equal the number in the most common morphology.}
In Figure \ref{fig11}, the H-similarity of late-type galaxies is lower than that of early-type galaxies. The more complex the structure of late-type galaxies, the more difficult it is to search for galaxies with high similarity. 
\COM{To address the limitations of insufficient data and lack of diversity, we propose a strategy centered on data augmentation using Variational Autoencoder (VAE; \cite{Kingma2013}) models. VAEs, renowned for their effectiveness in generating diverse images, provide a mechanism to adjust the variations of latent features specific to a galaxy, thereby enabling the generation of distinct yet similar images even for sparsely populated or unique galaxies. This unique capability of VAEs allows for the enrichment of our dataset across all morphologies, effectively overcoming the problem of data scarcity. By incorporating VAEs into our methodology, we anticipate an enhanced ability to accurately differentiate between various galaxy morphologies, even in scenarios of limited training data.}


Second, the compressed latent feature information of the galaxy images should be unbiased toward certain galaxy features. 
\COM{It is noted that late-type galaxies are occasionally misclassified as early-type galaxies, as seen in Figure \ref{fig18}. This misclassification arises predominantly due to the bias toward latent features that primarily represent overall color and shape, rather than detailed structural features like spiral arms and barred structures. To an experimental verification, we transformed galaxy images to grayscale, retrained our model, and confirmed subsequent classification changes.  Approximately 47$\%$ of galaxies initially misclassified from late-type to early-type were identified as late-type galaxies using grayscale images, suggesting a significant color effect. However, the remaining 53$\%$ were still classified as early-type. This outcome highlighted another limitation of our current model - its capacity to reconstruct delicate and faint structural details, which are crucial for precise morphological categorization. The Sa-type spiral galaxies, which are structurally close to the earlier types, clearly demonstrate these limitations. Even when the Sa-type was classified through a model that using gray scale image, about 67$\%$ were still not classified as late-type. To mitigate these biases and improve the accuracy of morphological classification, our future methodology needs to incorporate more comprehensive training data. Instead of relying solely on three-channel color data, the inclusion of more specific structural features of galaxies in the training data is paramount. In the next version of TSGICAS, we will include residual images (from galaxy-profile fitting model images) in the training dataset. This approach will enhance the representation of crucial structural features and reduce latent feature biases, thereby leading to a more accurate galaxy classification system.}



\section{Summary and Conclusions}\label{sec:summary}

We propose a novel and automatic galaxy morphological classification method employing the CAE and similarity indices. The CAE encoder converted the input galaxy images (256 × 256 × 3) into latent features (512) containing the morphological features. We then generated the morphological latent feature database. We compared the latent features of the database entries and the input galaxy images and calculated the similarity. Our method assigned the galaxy's morphology with the highest similarity in the database. The following briefly summarizes and concludes our key findings: 

1) \COM{Our CAE reproduces a highly similar image from an input galaxy image. The median and standard deviation of the mean pixel values for the residual images are 0.0181 and 0.003, respectively, which indicate a strong similarity between the input and reconstructed images. This suggests that the CAE effectively extracts the latent features, which include key morphological features of the input galaxy.}
 

2) \COM{The morphological \COMs{latent} feature database is constructed from galaxy images in GZ2, EFIGI, and NA10 catalogs. The entries from the EFIGI and NA10 catalogs exhibit a reliable agreement in morphological classifications, while the late-type galaxies in the GZ2 catalog are not as reliably classified. This reflects the inherent dependence of our morphological classification method, TSGICAS, on the quality of database classification. Understanding this limitation, it is important to recognize the potential ambiguities in visual classifications, as \cite{Cheng2021a} highlighted, where similar structural features could lead to varied or incorrect types. In the context of these findings, the exploration of unsupervised machine learning techniques, similar to the ones presented in \cite{Cheng2021a}, could contribute to improving the accuracy and reliability of the classification database, thereby enhancing the performance of the TSGICAS system.}

3) In most cases, even though the morphological types of MC samples and retrieved similar galaxies are different, the two galaxies can not be easily discriminated by visual inspection. Therefore, the TSGICAS suggests that the two galaxies may have the same morphological type. In addition, a catalog of well-classified galaxies should be used for accurate morphological classification. 

Since the encoder model of the CAE only extracts latent features from images, this classification method does not require our CAE re-learned for different morphological classification schemes for each catalog. Therefore, this method can automate the morphological classification of the tremendous amount of galaxies that will be observed in the next-generation telescopes.

The latent features of the CAE learned in this study showed a bias by color and overall shape of the galaxy (Section \ref{Misclass}; Figure \ref{fig18}). In future work, we aim to improve the model by expanding the training datasets by including the residual images representing faint structures. In addition, the correlation between the latent features and photometric parameters is examined, and their relative significance would enable accurate similarity calculations. Moreover, galaxies can be classified by clustering 512 latent features regardless of the existing morphological classification schemes. This can propose a new morphological classification scheme not presented in conventional morphological types. 

TSGICAS developed in this study can be used in various galaxy research and helps to easily and quickly find similar galaxies with a specific morphological type. Searching for similar images is more intuitive and efficient than the conventional classification method using photometric and/or spectroscopic parameters.
In particular, similar images of realistic simulation galaxies can be found in the observational domain. As a result, it is possible to compare the physical parameters of simulations and observations directly. Also, as the latent features imparted lower significance to the image noise, the output images exhibited a higher signal-to-noise ratio than the input images (Figure \ref{fig6}), thereby aiding in identifying the fainter and dimmer galaxy structures. 


\begin{acknowledgments}
We are grateful to the anonymous referee for helpful comments and suggestions that improved the clarity and quality of this paper. This work was supported by the National Research Foundation of Korea through grants NRF-2019R1I1A1A01061237 \& NRF-2022R1C1C2005539 (S.K.), NRF-2022R1I1A1A01054555 (Y.L.), NRF-2020R1I1A1A01052358 \& NRF-2021R1A2C1004117 (S.-I.H.), NRF- 2019R1A2C2086290 (H.-S.K.), NRF-2022R1A2C1007721 (S.C.R.), and NRF-2022M3K3A1093827 (H.S.).
\end{acknowledgments}

\bibliography{ref}{}

\begin{thebibliography}{}
\expandafter\ifx\csname natexlab\endcsname\relax\def\natexlab#1{#1}\fi
\providecommand{\url}[1]{\href{#1}{#1}}
\providecommand{\dodoi}[1]{doi:~\href{http://doi.org/#1}{\nolinkurl{#1}}}
\providecommand{\doeprint}[1]{\href{http://ascl.net/#1}{\nolinkurl{http://ascl.net/#1}}}
\providecommand{\doarXiv}[1]{\href{https://arxiv.org/abs/#1}{\nolinkurl{https://arxiv.org/abs/#1}}}

\bibitem[{{Abazajian} {et~al.}(2009){Abazajian}, {Adelman-McCarthy},
  {Ag{\"u}eros}, {Allam}, {Allende Prieto}, {An}, {Anderson}, {Anderson},
  {Annis}, {Bahcall}, {Bailer-Jones}, {Barentine}, {Bassett}, {Becker},
  {Beers}, {Bell}, {Belokurov}, {Berlind}, {Berman}, {Bernardi}, {Bickerton},
  {Bizyaev}, {Blakeslee}, {Blanton}, {Bochanski}, {Boroski}, {Brewington},
  {Brinchmann}, {Brinkmann}, {Brunner}, {Budav{\'a}ri}, {Carey}, {Carliles},
  {Carr}, {Castander}, {Cinabro}, {Connolly}, {Csabai}, {Cunha}, {Czarapata},
  {Davenport}, {de Haas}, {Dilday}, {Doi}, {Eisenstein}, {Evans}, {Evans},
  {Fan}, {Friedman}, {Frieman}, {Fukugita}, {G{\"a}nsicke}, {Gates},
  {Gillespie}, {Gilmore}, {Gonzalez}, {Gonzalez}, {Grebel}, {Gunn},
  {Gy{\"o}ry}, {Hall}, {Harding}, {Harris}, {Harvanek}, {Hawley}, {Hayes},
  {Heckman}, {Hendry}, {Hennessy}, {Hindsley}, {Hoblitt}, {Hogan}, {Hogg},
  {Holtzman}, {Hyde}, {Ichikawa}, {Ichikawa}, {Im}, {Ivezi{\'c}}, {Jester},
  {Jiang}, {Johnson}, {Jorgensen}, {Juri{\'c}}, {Kent}, {Kessler}, {Kleinman},
  {Knapp}, {Konishi}, {Kron}, {Krzesinski}, {Kuropatkin}, {Lampeitl},
  {Lebedeva}, {Lee}, {Lee}, {French Leger}, {L{\'e}pine}, {Li}, {Lima}, {Lin},
  {Long}, {Loomis}, {Loveday}, {Lupton}, {Magnier}, {Malanushenko},
  {Malanushenko}, {Mandelbaum}, {Margon}, {Marriner}, {Mart{\'\i}nez-Delgado},
  {Matsubara}, {McGehee}, {McKay}, {Meiksin}, {Morrison}, {Mullally}, {Munn},
  {Murphy}, {Nash}, {Nebot}, {Neilsen}, {Newberg}, {Newman}, {Nichol},
  {Nicinski}, {Nieto-Santisteban}, {Nitta}, {Okamura}, {Oravetz}, {Ostriker},
  {Owen}, {Padmanabhan}, {Pan}, {Park}, {Pauls}, {Peoples}, {Percival}, {Pier},
  {Pope}, {Pourbaix}, {Price}, {Purger}, {Quinn}, {Raddick}, {Re Fiorentin},
  {Richards}, {Richmond}, {Riess}, {Rix}, {Rockosi}, {Sako}, {Schlegel},
  {Schneider}, {Scholz}, {Schreiber}, {Schwope}, {Seljak}, {Sesar}, {Sheldon},
  {Shimasaku}, {Sibley}, {Simmons}, {Sivarani}, {Allyn Smith}, {Smith},
  {Smol{\v{c}}i{\'c}}, {Snedden}, {Stebbins}, {Steinmetz}, {Stoughton},
  {Strauss}, {SubbaRao}, {Suto}, {Szalay}, {Szapudi}, {Szkody}, {Tanaka},
  {Tegmark}, {Teodoro}, {Thakar}, {Tremonti}, {Tucker}, {Uomoto}, {Vanden
  Berk}, {Vandenberg}, {Vidrih}, {Vogeley}, {Voges}, {Vogt}, {Wadadekar},
  {Watters}, {Weinberg}, {West}, {White}, {Wilhite}, {Wonders}, {Yanny},
  {Yocum}, {York}, {Zehavi}, {Zibetti}, \& {Zucker}}]{Abazajian2009}
{Abazajian}, K.~N., {Adelman-McCarthy}, J.~K., {Ag{\"u}eros}, M.~A., {et~al.}
  2009, \apjs, 182, 543, \dodoi{10.1088/0067-0049/182/2/543}

\bibitem[{{Abraham} {et~al.}(2003){Abraham}, {van den Bergh}, \&
  {Nair}}]{Abraham2003}
{Abraham}, R.~G., {van den Bergh}, S., \& {Nair}, P. 2003, \apj, 588, 218,
  \dodoi{10.1086/373919}

\bibitem[{{Baillard} {et~al.}(2011){Baillard}, {Bertin}, {de Lapparent},
  {Fouqu{\'e}}, {Arnouts}, {Mellier}, {Pell{\'o}}, {Leborgne}, {Prugniel},
  {Makarov}, {Makarova}, {McCracken}, {Bijaoui}, \& {Tasca}}]{Baillard2011}
{Baillard}, A., {Bertin}, E., {de Lapparent}, V., {et~al.} 2011, \aap, 532,
  A74, \dodoi{10.1051/0004-6361/201016423}

\bibitem[{{Ball} {et~al.}(2004){Ball}, {Loveday}, {Fukugita}, {Nakamura},
  {Okamura}, {Brinkmann}, \& {Brunner}}]{Ball2004}
{Ball}, N.~M., {Loveday}, J., {Fukugita}, M., {et~al.} 2004, \mnras, 348, 1038,
  \dodoi{10.1111/j.1365-2966.2004.07429.x}

\bibitem[{{Beck} {et~al.}(2018){Beck}, {Scarlata}, {Fortson}, {Lintott},
  {Simmons}, {Galloway}, {Willett}, {Dickinson}, {Masters}, {Marshall}, \&
  {Wright}}]{Beck2018}
{Beck}, M.~R., {Scarlata}, C., {Fortson}, L.~F., {et~al.} 2018, \mnras, 476,
  5516, \dodoi{10.1093/mnras/sty503}

\bibitem[{{Bradley} {et~al.}(2016){Bradley}, {Sipocz}, {Robitaille},
  {Tollerud}, {Deil}, {Vin{\'\i}cius}, {Barbary}, {G{\"u}nther}, {Bostroem},
  {Droettboom}, {Bray}, {Bratholm}, {Pickering}, {Craig}, {Pascual}, {Greco},
  {Donath}, {Kerzendorf}, {Littlefair}, {Barentsen}, {D'Eugenio}, \&
  {Weaver}}]{Bradley2016}
{Bradley}, L., {Sipocz}, B., {Robitaille}, T., {et~al.} 2016, {Photutils:
  Photometry tools}, Astrophysics Source Code Library, record ascl:1609.011.
\newblock \doeprint{1609.011}

\bibitem[{{Cavanagh} {et~al.}(2021){Cavanagh}, {Bekki}, \&
  {Groves}}]{Cavanagh2021}
{Cavanagh}, M.~K., {Bekki}, K., \& {Groves}, B.~A. 2021, \mnras, 506, 659,
  \dodoi{10.1093/mnras/stab1552}

\bibitem[{{Cheng} {et~al.}(2021{\natexlab{a}}){Cheng}, {Huertas-Company},
  {Conselice}, {Arag{\'o}n-Salamanca}, {Robertson}, \&
  {Ramachandra}}]{Cheng2021a}
{Cheng}, T.-Y., {Huertas-Company}, M., {Conselice}, C.~J., {et~al.}
  2021{\natexlab{a}}, \mnras, 503, 4446, \dodoi{10.1093/mnras/stab734}

\bibitem[{{Cheng} {et~al.}(2020{\natexlab{a}}){Cheng}, {Li}, {Conselice},
  {Arag{\'o}n-Salamanca}, {Dye}, \& {Metcalf}}]{Cheng2020A}
{Cheng}, T.-Y., {Li}, N., {Conselice}, C.~J., {et~al.} 2020{\natexlab{a}},
  \mnras, 494, 3750, \dodoi{10.1093/mnras/staa1015}

\bibitem[{{Cheng} {et~al.}(2020{\natexlab{b}}){Cheng}, {Conselice},
  {Arag{\'o}n-Salamanca}, {Li}, {Bluck}, {Hartley}, {Annis}, {Brooks}, {Doel},
  {Garc{\'\i}a-Bellido}, {James}, {Kuehn}, {Kuropatkin}, {Smith}, {Sobreira},
  \& {Tarle}}]{Cheng2020}
{Cheng}, T.-Y., {Conselice}, C.~J., {Arag{\'o}n-Salamanca}, A., {et~al.}
  2020{\natexlab{b}}, \mnras, 493, 4209, \dodoi{10.1093/mnras/staa501}

\bibitem[{{Cheng} {et~al.}(2021{\natexlab{b}}){Cheng}, {Conselice},
  {Arag{\'o}n-Salamanca}, {Aguena}, {Allam}, {Andrade-Oliveira}, {Annis},
  {Bluck}, {Brooks}, {Burke}, {Carrasco Kind}, {Carretero}, {Choi}, {Costanzi},
  {da Costa}, {Pereira}, {De Vicente}, {Diehl}, {Drlica-Wagner}, {Eckert},
  {Everett}, {Evrard}, {Ferrero}, {Fosalba}, {Frieman}, {Garc{\'\i}a-Bellido},
  {Gerdes}, {Giannantonio}, {Gruen}, {Gruendl}, {Gschwend}, {Gutierrez},
  {Hinton}, {Hollowood}, {Honscheid}, {James}, {Krause}, {Kuehn}, {Kuropatkin},
  {Lahav}, {Maia}, {March}, {Menanteau}, {Miquel}, {Morgan},
  {Paz-Chinch{\'o}n}, {Pieres}, {Plazas Malag{\'o}n}, {Roodman}, {Sanchez},
  {Scarpine}, {Serrano}, {Sevilla-Noarbe}, {Smith}, {Soares-Santos}, {Suchyta},
  {Swanson}, {Tarle}, {Thomas}, \& {To}}]{Cheng2021}
---. 2021{\natexlab{b}}, \mnras, 507, 4425, \dodoi{10.1093/mnras/stab2142}

\bibitem[{{Cheng} {et~al.}(2023){Cheng}, {Dom{\'\i}nguez S{\'a}nchez},
  {Vega-Ferrero}, {Conselice}, {Siudek}, {Arag{\'o}n-Salamanca}, {Bernardi},
  {Cooke}, {Ferreira}, {Huertas-Company}, {Krywult}, {Palmese}, {Pieres},
  {Plazas Malag{\'o}n}, {Carnero Rosell}, {Gruen}, {Thomas}, {Bacon}, {Brooks},
  {James}, {Hollowood}, {Friedel}, {Suchyta}, {Sanchez}, {Menanteau},
  {Paz-Chinch{\'o}n}, {Gutierrez}, {Tarle}, {Sevilla-Noarbe}, {Ferrero},
  {Annis}, {Frieman}, {Garc{\'\i}a-Bellido}, {Mena-Fern{\'a}ndez}, {Honscheid},
  {Kuehn}, {da Costa}, {Gatti}, {Raveri}, {Pereira}, {Rodriguez-Monroy},
  {Smith}, {Carrasco Kind}, {Aguena}, {Swanson}, {Weaverdyck}, {Doel},
  {Miquel}, {Ogando}, {Gruendl}, {Allam}, {Hinton}, {Dodelson}, {Bocquet},
  {Desai}, {Everett}, \& {Scarpine}}]{Cheng2023}
{Cheng}, T.~Y., {Dom{\'\i}nguez S{\'a}nchez}, H., {Vega-Ferrero}, J., {et~al.}
  2023, \mnras, 518, 2794, \dodoi{10.1093/mnras/stac3228}

\bibitem[{{Conselice}(2003)}]{Conselice2003}
{Conselice}, C.~J. 2003, \apjs, 147, 1, \dodoi{10.1086/375001}

\bibitem[{{de la Calleja} \& {Fuentes}(2004)}]{DeLaCalleja2004}
{de la Calleja}, J., \& {Fuentes}, O. 2004, \mnras, 349, 87,
  \dodoi{10.1111/j.1365-2966.2004.07442.x}

\bibitem[{{de Vaucouleurs}(1959)}]{Vaucouleurs1959}
{de Vaucouleurs}, G. 1959, Handbuch der Physik, 53, 275,
  \dodoi{10.1007/978-3-642-45932-0_7}

\bibitem[{{de Vaucouleurs}(1963)}]{deVaucouleurs1963}
---. 1963, \apjs, 8, 31, \dodoi{10.1086/190084}

\bibitem[{Deng {et~al.}(2009)Deng, Dong, Socher, Li, Li, \& Fei-Fei}]{Deng09}
Deng, J., Dong, W., Socher, R., {et~al.} 2009, in 2009 IEEE Conference on
  Computer Vision and Pattern Recognition (CVPR), Vol.~0, 248--255,
  \dodoi{10.1109/CVPR.2009.5206848}

\bibitem[{{Dieleman} {et~al.}(2015){Dieleman}, {Willett}, \&
  {Dambre}}]{Dieleman2015}
{Dieleman}, S., {Willett}, K.~W., \& {Dambre}, J. 2015, \mnras, 450, 1441,
  \dodoi{10.1093/mnras/stv632}

\bibitem[{{Dubath} {et~al.}(2011){Dubath}, {Rimoldini}, {S{\"u}veges},
  {Blomme}, {L{\'o}pez}, {Sarro}, {De Ridder}, {Cuypers}, {Guy}, {Lecoeur},
  {Nienartowicz}, {Jan}, {Beck}, {Mowlavi}, {De Cat}, {Lebzelter}, \&
  {Eyer}}]{Dubath2011}
{Dubath}, P., {Rimoldini}, L., {S{\"u}veges}, M., {et~al.} 2011, \mnras, 414,
  2602, \dodoi{10.1111/j.1365-2966.2011.18575.x}

\bibitem[{{Falcon} {et~al.}(2020){Falcon}, {Borovec}, {W{\"a}lchli}, {Eggert},
  {Schock}, {Jordan}, {Skafte}, {Ir1dXD}, {Bereznyuk}, {Harris}, {Murrell},
  {Yu}, {Pr{\ae}sius}, {Addair}, {Zhong}, {Lipin}, {Uchida}, {Bapat},
  {Schr{\"o}ter}, {Dayma}, {Karnachev}, {Kulkarni}, {Komatsu}, {B},
  {SCHIRATTI}, {Mary}, {Byrne}, {Eyzaguirre}, {cinjon}, \&
  {Bakhtin}}]{Falcon2020}
{Falcon}, W., {Borovec}, J., {W{\"a}lchli}, A., {et~al.} 2020,
  {PyTorchLightning/pytorch-lightning: 0.7.6 release}, 0.7.6, Zenodo,  Zenodo,
  \dodoi{10.5281/zenodo.3828935}

\bibitem[{{Fukushima}(1975)}]{Fukushima1975}
{Fukushima}, K. 1975, Biological Cybernetics, 20, 121

\bibitem[{{Fukushima}(1980)}]{Fukushima1980}
---. 1980, Biological Cybernetics, 36, 193, \dodoi{10.1007/BF00344251}

\bibitem[{{Fukushima} {et~al.}(1983){Fukushima}, {Miyake}, \&
  {Ito}}]{Fukushima1983}
{Fukushima}, K., {Miyake}, S., \& {Ito}, T. 1983, IEEE Transactions on Systems,
  Man, and Cybernetics, SMC-13, 826, \dodoi{10.1109/TSMC.1983.6313076}

\bibitem[{{Gauci} {et~al.}(2010){Gauci}, {Zarb Adami}, \& {Abela}}]{Gauci2010}
{Gauci}, A., {Zarb Adami}, K., \& {Abela}, J. 2010, arXiv e-prints,
  arXiv:1005.0390.
\newblock \doarXiv{1005.0390}

\bibitem[{{Ghosh} {et~al.}(2020){Ghosh}, {Urry}, {Wang}, {Schawinski}, {Turp},
  \& {Powell}}]{Ghosh2020}
{Ghosh}, A., {Urry}, C.~M., {Wang}, Z., {et~al.} 2020, \apj, 895, 112,
  \dodoi{10.3847/1538-4357/ab8a47}

\bibitem[{{He} {et~al.}(2015){He}, {Zhang}, {Ren}, \& {Sun}}]{He2015}
{He}, K., {Zhang}, X., {Ren}, S., \& {Sun}, J. 2015, arXiv e-prints,
  arXiv:1512.03385.
\newblock \doarXiv{1512.03385}

\bibitem[{{Hocking} {et~al.}(2018){Hocking}, {Geach}, {Sun}, \&
  {Davey}}]{Hocking2018}
{Hocking}, A., {Geach}, J.~E., {Sun}, Y., \& {Davey}, N. 2018, \mnras, 473,
  1108, \dodoi{10.1093/mnras/stx2351}

\bibitem[{{Hubble}(1926)}]{Hubble1926}
{Hubble}, E.~P. 1926, \apj, 64, 321, \dodoi{10.1086/143018}

\bibitem[{{Huertas-Company} {et~al.}(2011){Huertas-Company}, {Aguerri},
  {Bernardi}, {Mei}, \& {S{\'a}nchez Almeida}}]{Huertas-Company2011}
{Huertas-Company}, M., {Aguerri}, J.~A.~L., {Bernardi}, M., {Mei}, S., \&
  {S{\'a}nchez Almeida}, J. 2011, \aap, 525, A157,
  \dodoi{10.1051/0004-6361/201015735}

\bibitem[{{Huertas-Company} {et~al.}(2008){Huertas-Company}, {Rouan}, {Tasca},
  {Soucail}, \& {Le F{\`e}vre}}]{Huertas-Company2008}
{Huertas-Company}, M., {Rouan}, D., {Tasca}, L., {Soucail}, G., \& {Le
  F{\`e}vre}, O. 2008, \aap, 478, 971, \dodoi{10.1051/0004-6361:20078625}

\bibitem[{{Huertas-Company} {et~al.}(2009){Huertas-Company}, {Tasca}, {Rouan},
  {Pelat}, {Kneib}, {Le F{\`e}vre}, {Capak}, {Kartaltepe}, {Koekemoer},
  {McCracken}, {Salvato}, {Sanders}, \& {Willott}}]{Huertas-Company2009}
{Huertas-Company}, M., {Tasca}, L., {Rouan}, D., {et~al.} 2009, \aap, 497, 743,
  \dodoi{10.1051/0004-6361/200811255}

\bibitem[{{Huertas-Company} {et~al.}(2015){Huertas-Company}, {Gravet},
  {Cabrera-Vives}, {P{\'e}rez-Gonz{\'a}lez}, {Kartaltepe}, {Barro}, {Bernardi},
  {Mei}, {Shankar}, {Dimauro}, {Bell}, {Kocevski}, {Koo}, {Faber}, \&
  {Mcintosh}}]{Huertas-Company2015}
{Huertas-Company}, M., {Gravet}, R., {Cabrera-Vives}, G., {et~al.} 2015, \apjs,
  221, 8, \dodoi{10.1088/0067-0049/221/1/8}

\bibitem[{{Huertas-Company} {et~al.}(2018){Huertas-Company}, {Primack},
  {Dekel}, {Koo}, {Lapiner}, {Ceverino}, {Simons}, {Snyder}, {Bernardi},
  {Chen}, {Dom{\'\i}nguez-S{\'a}nchez}, {Lee}, {Margalef-Bentabol}, \&
  {Tuccillo}}]{Huertas-Company2018}
{Huertas-Company}, M., {Primack}, J.~R., {Dekel}, A., {et~al.} 2018, \apj, 858,
  114, \dodoi{10.3847/1538-4357/aabfed}

\bibitem[{{Khalifa} {et~al.}(2017){Khalifa}, {Taha}, {Hassanien}, \&
  {Selim}}]{Khalifa2017}
{Khalifa}, N. E.~M., {Taha}, M. H.~N., {Hassanien}, A.~E., \& {Selim}, I.~M.
  2017, arXiv e-prints, arXiv:1709.02245.
\newblock \doarXiv{1709.02245}

\bibitem[{{Kingma} \& {Welling}(2013)}]{Kingma2013}
{Kingma}, D.~P., \& {Welling}, M. 2013, arXiv e-prints, arXiv:1312.6114,
  \dodoi{10.48550/arXiv.1312.6114}

\bibitem[{{Lahav} {et~al.}(1996){Lahav}, {Naim}, {Sodr{\'e}}, \&
  {Storrie-Lombardi}}]{Lahav1996}
{Lahav}, O., {Naim}, A., {Sodr{\'e}}, L., J., \& {Storrie-Lombardi}, M.~C.
  1996, \mnras, 283, 207, \dodoi{10.1093/mnras/283.1.207}

\bibitem[{{Law} {et~al.}(2007){Law}, {Steidel}, {Erb}, {Pettini}, {Reddy},
  {Shapley}, {Adelberger}, \& {Simenc}}]{Law2007}
{Law}, D.~R., {Steidel}, C.~C., {Erb}, D.~K., {et~al.} 2007, \apj, 656, 1,
  \dodoi{10.1086/510357}

\bibitem[{{Lintott} {et~al.}(2011){Lintott}, {Schawinski}, {Bamford}, {Slosar},
  {Land}, {Thomas}, {Edmondson}, {Masters}, {Nichol}, {Raddick}, {Szalay},
  {Andreescu}, {Murray}, \& {Vandenberg}}]{Lintott2011}
{Lintott}, C., {Schawinski}, K., {Bamford}, S., {et~al.} 2011, \mnras, 410,
  166, \dodoi{10.1111/j.1365-2966.2010.17432.x}

\bibitem[{{Lintott} {et~al.}(2008){Lintott}, {Schawinski}, {Slosar}, {Land},
  {Bamford}, {Thomas}, {Raddick}, {Nichol}, {Szalay}, {Andreescu}, {Murray}, \&
  {Vandenberg}}]{Lintott2008}
{Lintott}, C.~J., {Schawinski}, K., {Slosar}, A., {et~al.} 2008, \mnras, 389,
  1179, \dodoi{10.1111/j.1365-2966.2008.13689.x}

\bibitem[{{Lotz} {et~al.}(2004){Lotz}, {Primack}, \& {Madau}}]{Lotz2004}
{Lotz}, J.~M., {Primack}, J., \& {Madau}, P. 2004, \aj, 128, 163,
  \dodoi{10.1086/421849}

\bibitem[{{Maehoenen} \& {Hakala}(1995)}]{Maehoenen1995}
{Maehoenen}, P.~H., \& {Hakala}, P.~J. 1995, \apjl, 452, L77,
  \dodoi{10.1086/309697}

\bibitem[{{Naim} {et~al.}(1995){Naim}, {Lahav}, {Sodre}, \&
  {Storrie-Lombardi}}]{Naim1995}
{Naim}, A., {Lahav}, O., {Sodre}, L., J., \& {Storrie-Lombardi}, M.~C. 1995,
  \mnras, 275, 567, \dodoi{10.1093/mnras/275.3.567}

\bibitem[{{Nair} \& {Abraham}(2010)}]{Nair2010}
{Nair}, P.~B., \& {Abraham}, R.~G. 2010, \apjs, 186, 427,
  \dodoi{10.1088/0067-0049/186/2/427}

\bibitem[{{Odewahn} {et~al.}(1992){Odewahn}, {Stockwell}, {Pennington},
  {Humphreys}, \& {Zumach}}]{Odewahn1992}
{Odewahn}, S.~C., {Stockwell}, E.~B., {Pennington}, R.~L., {Humphreys}, R.~M.,
  \& {Zumach}, W.~A. 1992, \aj, 103, 318, \dodoi{10.1086/116063}

\bibitem[{{Paszke} {et~al.}(2019){Paszke}, {Gross}, {Massa}, {Lerer},
  {Bradbury}, {Chanan}, {Killeen}, {Lin}, {Gimelshein}, {Antiga}, {Desmaison},
  {K{\"o}pf}, {Yang}, {DeVito}, {Raison}, {Tejani}, {Chilamkurthy}, {Steiner},
  {Fang}, {Bai}, \& {Chintala}}]{Paszke2019}
{Paszke}, A., {Gross}, S., {Massa}, F., {et~al.} 2019, arXiv e-prints,
  arXiv:1912.01703.
\newblock \doarXiv{1912.01703}

\bibitem[{{Polsterer} {et~al.}(2012){Polsterer}, {Gieseke}, \&
  {Kramer}}]{Polsterer2012}
{Polsterer}, K.~L., {Gieseke}, F., \& {Kramer}, O. 2012, in Astronomical
  Society of the Pacific Conference Series, Vol. 461, Astronomical Data
  Analysis Software and Systems XXI, ed. P.~{Ballester}, D.~{Egret}, \&
  N.~P.~F. {Lorente}, 561

\bibitem[{{Ralph} {et~al.}(2019){Ralph}, {Norris}, {Fang}, {Park}, {Galvin},
  {Alger}, {Andernach}, {Lintott}, {Rudnick}, {Shabala}, \& {Wong}}]{Ralph2019}
{Ralph}, N.~O., {Norris}, R.~P., {Fang}, G., {et~al.} 2019, \pasp, 131, 108011,
  \dodoi{10.1088/1538-3873/ab213d}

\bibitem[{{Rumelhart} {et~al.}(1986){Rumelhart}, {Hinton}, \&
  {Williams}}]{Rumelhart1986}
{Rumelhart}, D.~E., {Hinton}, G.~E., \& {Williams}, R.~J. 1986, \nat, 323, 533,
  \dodoi{10.1038/323533a0}

\bibitem[{{Shamir}(2009)}]{Shamir2009}
{Shamir}, L. 2009, \mnras, 399, 1367, \dodoi{10.1111/j.1365-2966.2009.15366.x}

\bibitem[{{Sreejith} {et~al.}(2018){Sreejith}, {Pereverzyev}, {Kelvin},
  {Marleau}, {Haltmeier}, {Ebner}, {Bland-Hawthorn}, {Driver}, {Graham},
  {Holwerda}, {Hopkins}, {Liske}, {Loveday}, {Moffett}, {Pimbblet}, {Taylor},
  {Wang}, \& {Wright}}]{Sreejith2018}
{Sreejith}, S., {Pereverzyev}, Sergiy, J., {Kelvin}, L.~S., {et~al.} 2018,
  \mnras, 474, 5232, \dodoi{10.1093/mnras/stx2976}

\bibitem[{{Storey-Fisher} {et~al.}(2021){Storey-Fisher}, {Huertas-Company},
  {Ramachandra}, {Lanusse}, {Leauthaud}, {Luo}, {Huang}, \&
  {Prochaska}}]{Storey-Fisher2021}
{Storey-Fisher}, K., {Huertas-Company}, M., {Ramachandra}, N., {et~al.} 2021,
  \mnras, 508, 2946, \dodoi{10.1093/mnras/stab2589}

\bibitem[{{Strauss} {et~al.}(2002){Strauss}, {Weinberg}, {Lupton}, {Narayanan},
  {Annis}, {Bernardi}, {Blanton}, {Burles}, {Connolly}, {Dalcanton}, {Doi},
  {Eisenstein}, {Frieman}, {Fukugita}, {Gunn}, {Ivezi{\'c}}, {Kent}, {Kim},
  {Knapp}, {Kron}, {Munn}, {Newberg}, {Nichol}, {Okamura}, {Quinn}, {Richmond},
  {Schlegel}, {Shimasaku}, {SubbaRao}, {Szalay}, {Vanden Berk}, {Vogeley},
  {Yanny}, {Yasuda}, {York}, \& {Zehavi}}]{Strauss2002}
{Strauss}, M.~A., {Weinberg}, D.~H., {Lupton}, R.~H., {et~al.} 2002, \aj, 124,
  1810, \dodoi{10.1086/342343}

\bibitem[{{Vega-Ferrero} {et~al.}(2021){Vega-Ferrero}, {Dom{\'\i}nguez
  S{\'a}nchez}, {Bernardi}, {Huertas-Company}, {Morgan}, {Margalef}, {Aguena},
  {Allam}, {Annis}, {Avila}, {Bacon}, {Bertin}, {Brooks}, {Carnero Rosell},
  {Carrasco Kind}, {Carretero}, {Choi}, {Conselice}, {Costanzi}, {da Costa},
  {Pereira}, {De Vicente}, {Desai}, {Ferrero}, {Fosalba}, {Frieman},
  {Garc{\'\i}a-Bellido}, {Gruen}, {Gruendl}, {Gschwend}, {Gutierrez},
  {Hartley}, {Hinton}, {Hollowood}, {Honscheid}, {Hoyle}, {Jarvis}, {Kim},
  {Kuehn}, {Kuropatkin}, {Lima}, {Maia}, {Menanteau}, {Miquel}, {Ogando},
  {Palmese}, {Paz-Chinch{\'o}n}, {Plazas}, {Romer}, {Sanchez}, {Scarpine},
  {Schubnell}, {Serrano}, {Sevilla-Noarbe}, {Smith}, {Suchyta}, {Swanson},
  {Tarle}, {Tarsitano}, {To}, {Tucker}, {Varga}, \&
  {Wilkinson}}]{Vega-Ferrero2021}
{Vega-Ferrero}, J., {Dom{\'\i}nguez S{\'a}nchez}, H., {Bernardi}, M., {et~al.}
  2021, \mnras, 506, 1927, \dodoi{10.1093/mnras/stab594}

\bibitem[{{Walmsley} {et~al.}(2020){Walmsley}, {Smith}, {Lintott}, {Gal},
  {Bamford}, {Dickinson}, {Fortson}, {Kruk}, {Masters}, {Scarlata}, {Simmons},
  {Smethurst}, \& {Wright}}]{Walmsley2020}
{Walmsley}, M., {Smith}, L., {Lintott}, C., {et~al.} 2020, \mnras, 491, 1554,
  \dodoi{10.1093/mnras/stz2816}

\bibitem[{{Walmsley} {et~al.}(2022){Walmsley}, {Scaife}, {Lintott}, {Lochner},
  {Etsebeth}, {G{\'e}ron}, {Dickinson}, {Fortson}, {Kruk}, {Masters}, {Mantha},
  \& {Simmons}}]{Walmsley2022}
{Walmsley}, M., {Scaife}, A. M.~M., {Lintott}, C., {et~al.} 2022, \mnras, 513,
  1581, \dodoi{10.1093/mnras/stac525}

\bibitem[{{Weir} {et~al.}(1995){Weir}, {Fayyad}, \& {Djorgovski}}]{Weir1995}
{Weir}, N., {Fayyad}, U.~M., \& {Djorgovski}, S. 1995, \aj, 109, 2401,
  \dodoi{10.1086/117459}

\bibitem[{{Willett} {et~al.}(2013){Willett}, {Lintott}, {Bamford}, {Masters},
  {Simmons}, {Casteels}, {Edmondson}, {Fortson}, {Kaviraj}, {Keel}, {Melvin},
  {Nichol}, {Raddick}, {Schawinski}, {Simpson}, {Skibba}, {Smith}, \&
  {Thomas}}]{Willett2013}
{Willett}, K.~W., {Lintott}, C.~J., {Bamford}, S.~P., {et~al.} 2013, \mnras,
  435, 2835, \dodoi{10.1093/mnras/stt1458}

\bibitem[{{York} {et~al.}(2000){York}, {Adelman}, {Anderson}, {Anderson},
  {Annis}, {Bahcall}, {Bakken}, {Barkhouser}, {Bastian}, {Berman}, {Boroski},
  {Bracker}, {Briegel}, {Briggs}, {Brinkmann}, {Brunner}, {Burles}, {Carey},
  {Carr}, {Castander}, {Chen}, {Colestock}, {Connolly}, {Crocker}, {Csabai},
  {Czarapata}, {Davis}, {Doi}, {Dombeck}, {Eisenstein}, {Ellman}, {Elms},
  {Evans}, {Fan}, {Federwitz}, {Fiscelli}, {Friedman}, {Frieman}, {Fukugita},
  {Gillespie}, {Gunn}, {Gurbani}, {de Haas}, {Haldeman}, {Harris}, {Hayes},
  {Heckman}, {Hennessy}, {Hindsley}, {Holm}, {Holmgren}, {Huang}, {Hull},
  {Husby}, {Ichikawa}, {Ichikawa}, {Ivezi{\'c}}, {Kent}, {Kim}, {Kinney},
  {Klaene}, {Kleinman}, {Kleinman}, {Knapp}, {Korienek}, {Kron}, {Kunszt},
  {Lamb}, {Lee}, {Leger}, {Limmongkol}, {Lindenmeyer}, {Long}, {Loomis},
  {Loveday}, {Lucinio}, {Lupton}, {MacKinnon}, {Mannery}, {Mantsch}, {Margon},
  {McGehee}, {McKay}, {Meiksin}, {Merelli}, {Monet}, {Munn}, {Narayanan},
  {Nash}, {Neilsen}, {Neswold}, {Newberg}, {Nichol}, {Nicinski}, {Nonino},
  {Okada}, {Okamura}, {Ostriker}, {Owen}, {Pauls}, {Peoples}, {Peterson},
  {Petravick}, {Pier}, {Pope}, {Pordes}, {Prosapio}, {Rechenmacher}, {Quinn},
  {Richards}, {Richmond}, {Rivetta}, {Rockosi}, {Ruthmansdorfer}, {Sandford},
  {Schlegel}, {Schneider}, {Sekiguchi}, {Sergey}, {Shimasaku}, {Siegmund},
  {Smee}, {Smith}, {Snedden}, {Stone}, {Stoughton}, {Strauss}, {Stubbs},
  {SubbaRao}, {Szalay}, {Szapudi}, {Szokoly}, {Thakar}, {Tremonti}, {Tucker},
  {Uomoto}, {Vanden Berk}, {Vogeley}, {Waddell}, {Wang}, {Watanabe},
  {Weinberg}, {Yanny}, {Yasuda}, \& {SDSS Collaboration}}]{York2000}
{York}, D.~G., {Adelman}, J., {Anderson}, John~E., J., {et~al.} 2000, \aj, 120,
  1579, \dodoi{10.1086/301513}

\bibitem[{{Zeiler} \& {Fergus}(2013)}]{Zeiler2013}
{Zeiler}, M.~D., \& {Fergus}, R. 2013, arXiv e-prints, arXiv:1311.2901.
\newblock \doarXiv{1311.2901}

\end{thebibliography}
\bibliographystyle{aasjournal}



\end{document}